\documentclass{aa}

\usepackage[T1]{fontenc}
\usepackage[utf8]{inputenc}

\usepackage{graphicx}
\graphicspath{{./}{Figures/}{Plots/}}

\usepackage[pdfpagelabels=false]{hyperref}

\usepackage[dvipsnames]{xcolor}
\usepackage{xspace}
\usepackage{soul}
\usepackage{booktabs}
\usepackage{adjustbox}

\usepackage{amsmath}
\usepackage{commath}
\usepackage{upgreek}
\usepackage{siunitx}
\usepackage{nth}
\usepackage{newtxtext}
\usepackage{newtxmath}

\usepackage[capitalise, noabbrev]{cleveref}
\usepackage{enumitem}
\usepackage{csvsimple}
\usepackage{multirow}
\usepackage{array}
\usepackage{placeins}

\crefname{figure}{Fig.}{Figs.}
\crefname{equation}{equation}{equations}

\bibpunct[]{(}{)}{;}{a}{}{,}
\defcitealias{2025Natur.639..897W}{W25}

\DeclareRobustCommand{\VAN}[3]{#2}
\let\VANthebibliography\thebibliography
\def\thebibliography{\DeclareRobustCommand{\VAN}[3]{##3}\VANthebibliography}

\newcommand{\orcid}[2]{\href{http://orcid.org/#2}{#1}}
\newcommand{\orcidsymb}[2]{#1\href{http://orcid.org/#2}{\adjustbox{trim={-.15\width} {0\height} {-.15\width} {0\height},clip}{\includegraphics[height=10pt]{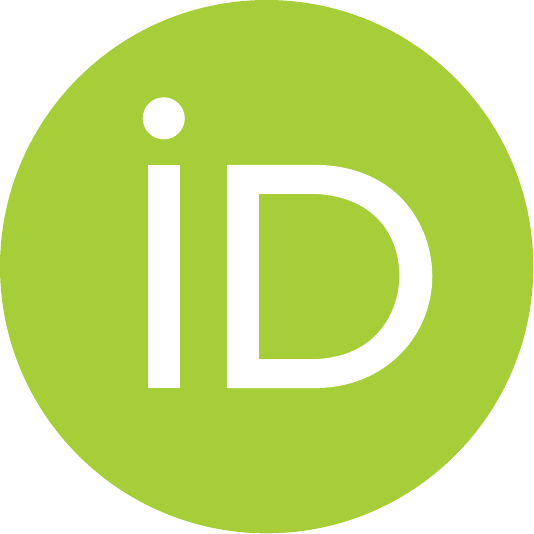}}}}

\newcommand{\ssim}{\sim \!}

\newcommand{\Lymana}{{Lyman-\ensuremath{\upalpha}}\xspace}
\newcommand{\Lymanatext}{Lyman-α}
\newcommand{\Lya}{{Ly\ensuremath{\upalpha}}\xspace}
\newcommand{\Lyatext}{Lyα}
\newcommand{\HI}{\hbox{H\,{\sc i}}\xspace}

\newcommand{\CIV}{\hbox{C\,{\sc iv}}\xspace}

\newcommand{\OII}{\hbox{[O\,{\sc ii}]}\xspace}
\newcommand{\OIII}{\hbox{[O\,{\sc iii}]}\xspace}

\newcommand{\Halpha}{\ensuremath{\mathrm{H}\upalpha}\xspace}
\newcommand{\Hbeta}{\ensuremath{\mathrm{H}\upbeta}\xspace}

\newcommand{\Paalpha}{\ensuremath{\mathrm{Pa}\upalpha}\xspace}

\newcommand{\JGSzthirteenLA}{GS-z13-1-LA\xspace}

\hypersetup{
    unicode=true,
    final=true,
    plainpages=false,
    pdfstartview=FitV,
    pdftoolbar=false,
    pdfmenubar=true,
    bookmarksopen=true,
    bookmarksnumbered=true,
    breaklinks=true,
    linktocpage,
    colorlinks=true,
    linkcolor=Blue,
    urlcolor=Blue,
    citecolor=Blue,
    anchorcolor=green
}
\AtBeginDocument{
    \hypersetup{
        pdftitle={An OASIS of \Lymanatext\ within a neutral intergalactic desert},
        pdfauthor={J. Witstok et al.},
        pdfsubject={Reaffirmed line and blue continuum reveal efficient ionising agents at z = 13},
        pdfkeywords={{galaxies: high-redshift} -- {dark ages, reionization, first stars} -- {methods: observational}}
    }
}

\begin{document}

\title{An OASIS of \texorpdfstring{\Lymana}{\Lymanatext} within a neutral intergalactic desert}
\subtitle{Reaffirmed line and blue continuum reveal efficient ionising agents at $z = 13$}
\titlerunning{An OASIS of \texorpdfstring{\Lymana}{\Lymanatext} within a neutral intergalactic desert}

\author{
    \protect\orcidsymb{Joris~Witstok}{0000-0002-7595-121X}\inst{\ref{inst:DAWN}}\fnmsep\inst{\ref{inst:NBI}}\corrauth{joris.witstok@nbi.ku.dk}
    \and \protect\orcidsymb{Stefano~Carniani}{0000-0002-6719-380X}\inst{\ref{inst:SNS}} \email{stefano.carniani@sns.it} \and \protect\orcidsymb{Peter~Jakobsen}{0000-0002-6780-2441}\inst{\ref{inst:DAWN}}\fnmsep\inst{\ref{inst:NBI}} \email{pjakobsen@nbi.ku.dk} \and \protect\orcidsymb{Andrew~J.~Bunker}{0000-0002-8651-9879}\inst{\ref{inst:Oxford}} \email{andy.bunker@physics.ox.ac.uk} \and \protect\orcidsymb{Alex~J.~Cameron}{0000-0002-0450-7306}\inst{\ref{inst:DAWN}}\fnmsep\inst{\ref{inst:NBI}} \email{alex.cameron@nbi.ku.dk} \and \protect\orcidsymb{Francesco~D'Eugenio}{0000-0003-2388-8172}\inst{\ref{inst:Kavli}}\fnmsep\inst{\ref{inst:Cav}} \email{fd391@cam.ac.uk} \and \protect\orcidsymb{Kevin~Hainline}{0000-0003-4565-8239}\inst{\ref{inst:Steward}} \email{kevinhainline@arizona.edu} \and \protect\orcidsymb{Jakob~M.~Helton}{0000-0003-4337-6211}\inst{\ref{inst:Penn}} \email{jakobhelton@psu.edu} \and \protect\orcidsymb{Tobias~J.~Looser}{0000-0002-3642-2446}\inst{\ref{inst:CfA}} \email{tobias.looser@cfa.harvard.edu} \and \protect\orcidsymb{Pierluigi~Rinaldi}{0000-0002-5104-8245}\inst{\ref{inst:STScI}} \email{prinaldi@stsci.edu} \and \protect\orcidsymb{Brant~E.~Robertson}{0000-0002-4271-0364}\inst{\ref{inst:UCSC}} \email{brant@ucsc.edu} \and \protect\orcidsymb{William~M.~Baker}{0000-0003-0215-1104}\inst{\ref{inst:DARK}} \email{william.baker@nbi.ku.dk} \and \protect\orcidsymb{Stéphane~Charlot}{0000-0003-3458-2275}\inst{\ref{inst:IAP}} \email{charlot@iap.fr} \and \protect\orcidsymb{Benjamin~D.~Johnson}{0000-0002-9280-7594}\inst{\ref{inst:CfA}} \email{benjamin.johnson@cfa.harvard.edu} \and \protect\orcidsymb{Gareth~C.~Jones}{0000-0002-0267-9024}\inst{\ref{inst:Kavli}}\fnmsep\inst{\ref{inst:Cav}} \email{gj283@cam.ac.uk} \and \protect\orcidsymb{Nimisha~Kumari}{0000-0002-5320-2568}\inst{\ref{inst:AURA}} \email{kumari@stsci.edu} \and \protect\orcidsymb{Roberto~Maiolino}{0000-0002-4985-3819}\inst{\ref{inst:Kavli}}\fnmsep\inst{\ref{inst:Cav}}\fnmsep\inst{\ref{inst:UCL}} \email{rm665@cam.ac.uk} \and \protect\orcidsymb{Jan~Scholtz}{0000-0001-6010-6809}\inst{\ref{inst:Kavli}}\fnmsep\inst{\ref{inst:Cav}} \email{honzascholtz@gmail.com} \and \protect\orcidsymb{Sandro~Tacchella}{0000-0002-8224-4505}\inst{\ref{inst:Kavli}}\fnmsep\inst{\ref{inst:Cav}} \email{st578@cam.ac.uk} \and \protect\orcidsymb{Christopher~N.~A.~Willmer}{0000-0001-9262-9997}\inst{\ref{inst:Steward}} \email{cnaw@as.arizona.edu} \and \protect\orcidsymb{Chris~Willott}{0000-0002-4201-7367}\inst{\ref{inst:NRC}} \email{chris.willott@nrc.ca} \and \protect\orcidsymb{Zihao~Wu}{0000-0002-8876-5248}\inst{\ref{inst:CfA}} \email{zihao.wu@cfa.harvard.edu}
    }
\authorrunning{J.~Witstok et al.}

\institute{
    Cosmic Dawn Center (DAWN), Copenhagen, Denmark\label{inst:DAWN}
    \and Niels Bohr Institute, University of Copenhagen, Jagtvej 128, DK-2200, Copenhagen, Denmark\label{inst:NBI}
    \and Scuola Normale Superiore, Piazza dei Cavalieri 7, I-56126 Pisa, Italy\label{inst:SNS}
    \and Department of Physics, University of Oxford, Denys Wilkinson Building, Keble Road, Oxford OX1 3RH, UK\label{inst:Oxford}
    \and Kavli Institute for Cosmology, University of Cambridge, Madingley Road, Cambridge CB3 0HA, UK\label{inst:Kavli}
    \and Cavendish Laboratory, University of Cambridge, 19 JJ Thomson Avenue, Cambridge CB3 0HE, UK\label{inst:Cav}
    \and Steward Observatory, University of Arizona, 933 N. Cherry Ave., Tucson, AZ 85721, USA\label{inst:Steward}
    \and Department of Astronomy \& Astrophysics, The Pennsylvania State University, University Park, PA 16802, USA\label{inst:Penn}
    \and Center for Astrophysics $|$ Harvard \& Smithsonian, 60 Garden St., Cambridge, MA 02138, USA\label{inst:CfA}
    \and Space Telescope Science Institute, 3700 San Martin Drive, Baltimore, MD 21218, USA\label{inst:STScI}
    \and Department of Astronomy and Astrophysics University of California, Santa Cruz, 1156 High Street, Santa Cruz, CA 96054, USA\label{inst:UCSC}
    \and DARK, Niels Bohr Institute, University of Copenhagen, Jagtvej 155A, DK-2200 Copenhagen, Denmark\label{inst:DARK}
    \and Sorbonne Universit\'e, CNRS, UMR 7095, Institut d'Astrophysique de Paris, 98 bis bd Arago, 75014 Paris, France\label{inst:IAP}
    \and AURA for European Space Agency, Space Telescope Science Institute, 3700 San Martin Drive, Baltimore, MD 21210, USA\label{inst:AURA}
    \and Department of Physics and Astronomy, University College London, Gower Street, London WC1E 6BT, UK\label{inst:UCL}
    \and NRC Herzberg, 5071 West Saanich Rd, Victoria, BC V9E 2E7, Canada\label{inst:NRC}
}

\date{Received \today}

\abstract{
     Galaxy assembly was already well underway in the first $400 \, \mathrm{Myr}$ of cosmic time, as recently revealed by JWST. However, the contribution of these early galaxies to cosmic reionisation remains uncertain. Here we present new JWST/NIRSpec observations of \JGSzthirteenLA obtained as part of the OASIS and JADES programmes, whose combined deep ($56 \, \mathrm{h}$) NIRSpec/PRISM spectrum confirms the \Lymana line detection and blue UV continuum at redshift $z = 13.1$ presented in a previous work. The measured \Lymana emission (rest-frame equivalent width of $66_{-9}^{+10} \, \Angstrom$) and steep continuum slope ($\beta_\text{UV} \approx -3$) point towards \JGSzthirteenLA hosting a remarkably hot and powerful ionising source, and allow at most a modest contribution from the nebular continuum. The steep turnover of the continuum is markedly less pronounced in the new OASIS spectrum, suggesting a lower \HI column density covering the source (or OASIS-specific sightline). Combined, this implies that ionising photons may escape \JGSzthirteenLA at a sufficient rate to weaken the other, still undetected UV lines, and lead to the formation of a small ionised bubble ($R_\text{ion} \approx 0.2 \, \mathrm{pMpc}$), although the precise bubble size is dependent on model assumptions. A yet larger bubble could alleviate the required ionising photon production efficiency of \JGSzthirteenLA from $\xi_\mathrm{{ion}} \approx 10^{26.4} \, \mathrm{{Hz \, erg^{{-1}}}}$ down to $\approx 10^{25.9} \, \mathrm{{Hz \, erg^{{-1}}}}$, still extremely high but more readily reconcilable with stellar models. In turn, this would require a notable overdensity of galaxies with highly efficient ionising capabilities, a scenario for which tentative evidence is found in the form of $17$ nearby photometric candidates and one spectroscopically confirmed source, JADES-GS-z13-0. The new OASIS observations therefore confirm the overall picture of \JGSzthirteenLA as an early beacon of reionisation, providing compelling evidence for its start only $330 \, \mathrm{Myr}$ after the Big Bang.
}


\maketitle
\nolinenumbers

\section{Introduction}
\label{sec:Introduction}

Most of the baryons in the Universe are contained in the intergalactic medium \citep[IGM;][]{2009RvMP...81.1405M}. Only a few hundred million years after the Big Bang, neutral atomic hydrogen (\HI) within this vast gas reservoir started to be photoionised and heated as a result of the formation of the first astrophysical objects, marking the onset of cosmic reionisation \citep{2018PhR...780....1D}.

While future \HI $21 \, \mathrm{cm}$ experiments hold promise to directly trace \HI gas in the early Universe \citep{2022NatAs...6..984D}, the \HI\ \Lymana (\Lya) transition is currently the most potent probe of reionisation \citep[see][ for a review]{2020ARA&A..58..617O}. In its closing stages, corresponding to redshift $z \lesssim 6$, the multitude (`forest') of \Lya absorption features observed in quasar spectra is now able to accurately constrain the endpoint of reionisation, which is generally argued to be `late' \citep[persisting down to $z \approx 5.3$;][]{2022MNRAS.514...55B, 2022ApJ...932...76Z}. The \Lya forest even allows the timeline of reionisation to be reconstructed further back in time ($z > 6$) with exquisite precision, although these constraints are subject to model dependency \citep[e.g.][]{2022ApJ...933...59V, 2025PASA...42...49Q}.

Notwithstanding its late ending, there have also been indications of an early start to reionisation. Recent re-analyses of the cosmic microwave background (CMB) revealed that this could relieve tensions between baryonic acoustic oscillation and CMB measurements \citep[; though other tensions may arise as a result, cf. \citealt{2025ApJ...987L..29C}]{2025PhRvD.112d3541J, 2026PhRvL.136h1002S, 2026ApJ..1003....7B}. Deeper into the heart of the reionisation era when quasars become exceedingly rare, the steadily declining \Lya emission strength observed in galaxy spectra can instead be used as a robust, independent probe of an increasingly neutral IGM \citep{2026OJAp....960756N}. While statistical analyses of recent JWST observations up to $z \approx 14$ infer a steady increase with redshift in the global neutral hydrogen fraction \citep{2024ApJ...975..208T, 2024A&A...683A.238J, 2025ApJS..278...33K}, \Lya emitting galaxies (LAEs) have long been known to exist even at $z > 7$ \citep[e.g.][]{2012ApJ...744...83O, 2013Natur.502..524F, 2015ApJ...804L..30O, 2015ApJ...810L..12Z, 2016ApJ...823..143R, 2022MNRAS.511.6042E, 2024ApJ...970...50C, 2025arXiv251206072L}.

Despite the substantial \Lya scattering cross section expected from a predominantly neutral IGM, several of these LAEs exhibit surprisingly high equivalent width (EW) emission arising near resonance \citep[up to several $100 \, \Angstrom$ in the rest frame; e.g.][]{2023A&A...678A..68S, 2024A&A...684A..84S, 2024MNRAS.531.2701T, 2024ApJ...972...56T, 2025ApJ...988...26W, 2025ApJ...993..194N, 2026A&A...708A.102N}. Even if the bulk of the IGM is still neutral, these galaxies are expected to be located in rare ionised oases or `bubbles' with sizes of the order of $1 \, \text{physical Mpc}$ \citep[pMpc;][]{2024A&A...682A..40W, 2025MNRAS.536...27W, 2025ApJ...984...95R, 2026ApJ..1001..236C, 2026ApJ...997..102M}, at times challenging theoretical predictions \citep[e.g.][]{2024MNRAS.528.4872L, 2024MNRAS.531.2943N}. On the other hand, there is mounting evidence for strong damped \Lya (DLA) absorption \citep[$N_\text{\HI} > 10^{22} \, \mathrm{cm^{-2}}$;][]{2024Sci...384..890H, 2025A&A...693A..60H, 2025ApJ...987L...2H, 2024ApJ...976..160H, 2024A&A...689A.152D, 2026A&A...711A.298P}, which suggests the presence of large, neutral gas reservoirs surrounding galaxies in their early assembly stages, potentially inhibiting the escape of Lyman-continuum (LyC) radiation. While the presence of $z > 8$ LAEs\footnote{EGSY8p7 (CEERS-1019) at $z = 8.7$ \citep{2015ApJ...810L..12Z}; EGS-z910-44164 (CEERS-1029) at $z = 8.6$ \citep[; though see \citealt{2024ApJ...967...28N}]{2022ApJ...930..104L, 2023MNRAS.526.1657T}; the triplet of JADES sources GS-z8-0-LA at $z = 8.5$ \citep{2024MNRAS.531.2701T}, GN-z8-0-LA at $z = 8.3$, and GS-z8-1-LA at $z = 8.7$ \citep{2025MNRAS.536...27W, 2025MNRAS.536.2355J}; GN-z11 at $z = 10.6$ \citep{2023A&A...677A..88B, 2024A&A...687A.283S}; and \JGSzthirteenLA at $z \approx 13$ \citep{2025Natur.639..897W}.} suggests that at least a small fraction of the Universe was already ionised, it remains unclear how common such early bubbles are, and when they first formed.

From the JWST Advanced Deep Extragalactic Survey \citep[JADES;][]{2026ApJS..283....6E}, a major surprise came with the discovery of exceptionally strong \Lya emission in JADES-GS-z13-1-LA (\JGSzthirteenLA for short) at $z \approx 13$, which exhibited a rest-frame EW of at least $43 \, \Angstrom$ \citep[; \citetalias{2025Natur.639..897W} hereafter]{2025Natur.639..897W}. Although seemingly at odds with a late and rapid reionisation scenario \citep[e.g.][]{2020ApJ...892..109N, 2021ApJ...917L..37C}, conventional reionisation models are able to accommodate such an early LAE: \citet{2025MNRAS.538L..16Q} predicted a \Lya transmission of $\ssim 13\%$ and estimated a probability of up to $10\%$ of observing similarly strong \Lya emission as seen in \JGSzthirteenLA. Similarly, the model calibrated on JWST UV luminosity functions in \citet{2026arXiv260617571Z} predicts an incidence of ionised bubbles as inferred by \citetalias{2025Natur.639..897W} for \JGSzthirteenLA of approximately one per $75 \, \mathrm{arcmin^{2}}$ and per $\Delta z = 1$, which would yield $N \approx 2$ expected instances at $z = 10$--$14$ over the deep portion of NIRCam imaging in GOODS-S \citep{2025ApJS..281...50E, 2026arXiv260115954J}. In \citet{2026PASA...43...46C}, it was estimated that the observed \Lya properties of \JGSzthirteenLA imply the global neutral fraction of the Universe at $z = 13$ should be $x_\text{\HI} \lesssim 95\%$, and that there is a non-negligible probability for this source to host an active galactic nucleus (AGN).

Based on Near-Infrared Camera \citep[NIRCam;][]{2023PASP..135b8001R} imaging in the JADES Origins Field (JOF) programme \citep{2025ApJS..281...50E}, \JGSzthirteenLA was selected as the most robust candidate among the $z \gtrsim 11$ sample identified by \citet{2024ApJ...970...31R}, where it appeared as a very blue, spatially compact $z \approx 13$ galaxy. Spectroscopy from the Near-Infrared Spectrograph \citep[NIRSpec;][]{2022A&A...661A..80J} subsequently uncovered the luminous \Lya line in addition to confirming its very steep UV slope \citepalias[$\beta_\text{UV} \approx -3$;][]{2025Natur.639..897W}, the latter uncommon but not unique among the reionisation-era galaxy population \citep{2023MNRAS.520...14C, 2024MNRAS.531..997C, 2024MNRAS.529.4087T, 2026MNRAS.548ag808S, 2025ApJ...995...43A, 2025ApJ...988...86Y, 2025A&A...697A..90B, 2026ApJ..1001..234J, 2026A&A...711A.301M}. Simultaneously, it revealed a smooth \Lya break that could be consistent with a dominant two-photon ($2\gamma$) emission \citep{2024MNRAS.534..523C}, although a nebular-dominated continuum appears difficult to reconcile with the blue continuum \citep{2025OJAp....8E.104K}. Alternatively, the turnover of the continuum towards \Lya may be explained by a strong DLA absorption system \citepalias[$N_\text{\HI} \approx 10^{22.8} \, \mathrm{cm^{-2}}$;][]{2025Natur.639..897W}. Effectively unresolved in NIRCam imaging, \citetalias{2025Natur.639..897W} conservatively estimated its half-light radius in the rest-frame UV to be $r \lesssim 35 \, \mathrm{pc}$.

Despite the $18.7 \, \mathrm{h}$ depth of the initial NIRSpec dataset presented in \citetalias{2025Natur.639..897W}, the intrinsic faintness of \JGSzthirteenLA called for confirmation and more precise characterisation of the remarkable physical properties inferred from the continuum shape and \Lya line emission. Improved signal-to-noise ratio (SNR) could elucidate the origin of the continuum turnover and reveal other rest-frame UV lines to pin down the systemic redshift, thus shedding light on the remarkable efficiency of ionising photon production and escape in \JGSzthirteenLA as implied by the \Lya emission. In this work, we present new NIRSpec observations consisting of a $3\times$ deeper exposure in NIRSpec/PRISM obtained as part of the Observing All phases of StochastIc Star formation (OASIS) Cycle-2 JWST programme (ID 5997, PIs: T.~Looser \& F.~D'Eugenio; \citealt{2026arXiv261041489L, 2026arXiv261041490L}).

The outline of this work is as follows. In \cref{sec:Observations}, we discuss the observations underlying this work. Our results are presented in \cref{sec:Results} and discussed in \cref{sec:Discussion}, with \cref{sec:Conclusions} providing a summary. We adopt a flat $\Lambda$CDM cosmology throughout, with $H_0 = 67.4 \, \mathrm{km \, s^{-1} \, Mpc^{-1}}$, $\Omega_\text{m} = 0.315$, $\Omega_\text{b} = 0.0492$ based on the latest results of the Planck collaboration \citep{2020A&A...641A...6P}. On-sky separations of $1\arcsec$ and $1\arcmin$ correspond to $3.53 \, \text{physical kpc}$ (pkpc) and $0.212 \, \mathrm{pMpc}$, respectively. Magnitudes are in the AB system \citep{1983ApJ...266..713O}, emission-line wavelengths in vacuum, and EWs in the rest frame.

\section{Observations}
\label{sec:Observations}

\subsection{JWST observing programmes}
\label{ssec:JWST_observing_programmes}

The NIRCam and NIRSpec observations of \JGSzthirteenLA presented in \citet{2024ApJ...970...31R} and \citetalias{2025Natur.639..897W}, respectively, are associated with JWST guaranteed time observations (GTO) programme IDs (PIDs) 1180 (PI: D.~Eisenstein), 1210, and 1287 (PI: N.~L\"{u}tzgendorf), and with the JOF general observer (GO) programme \citep[PID 3215, PIs: D.~Eisenstein \& R.~Maiolino;][]{2025ApJS..281...50E}. Details on the JADES survey strategy and data reduction can be found in the survey overview \citep{2026ApJS..283....6E} and data release (DR\footnote{JADES high-level science products are made publicly available at \url{https://archive.stsci.edu/hlsp/jades} (NIRCam DRs) as well as at \url{https://jades.herts.ac.uk/} (NIRSpec DR4). The associated Mikulski Archive for Space Telescopes (MAST) DOI is \url{https://doi.org/10.17909/8tdj-8n28}.}) papers for NIRCam \citep{2023ApJS..269...16R, 2026arXiv260115954J, 2026arXiv260115956R} and/or NIRSpec \citep{2024A&A...690A.288B, 2025ApJS..277....4D, 2026MNRAS.549ag836C, 2026MNRAS.549ag939S}. Within JADES, \JGSzthirteenLA is associated with NIRCam ID 13731 and NIRSpec ID 20013731.

The NIRSpec observations presented in \citetalias{2025Natur.639..897W} consisted of $18.7 \, \mathrm{h}$ in the low-resolution PRISM/CLEAR mode ($R \approx 100$; `PRISM' hereafter) and $4.7 \, \mathrm{h}$ in the medium-resolution G235M/F170LP grating-filter combination ($R \approx 1000$). They were obtained between 10 and 11 January 2024 as part of the DEEP/JWST tier in JADES \citep[PID 1287; see e.g.][ for more details on the survey tiers]{2026MNRAS.549ag939S}. In this work, we combine these data with the successful repetition of one of three visits that originally failed due to the loss of the guide-star lock \citep[see also][]{2024Natur.633..318C}, which adds $9.3 \, \mathrm{h}$ in the PRISM and $2.3 \, \mathrm{h}$ in G235M. Additionally, we make use of the $28 \, \mathrm{h}$ NIRSpec/PRISM measurements obtained as part of OASIS (PID 5997) between 6 and 7 January 2025, in which \JGSzthirteenLA was placed on the micro-shutter array \citep[MSA;][]{2022A&A...661A..81F} as a supplementary, high-value target.\footnote{NIRSpec data of \JGSzthirteenLA taken as part of OASIS are publicly available at MAST under \url{https://doi.org/10.17909/ymh1-7h48}.} These observations were taken with the same exposure sequence as in PID 1287, though with a different MSA configuration and hence intra-shutter position, as shown in \cref{fig:Spectra_comparison}. New total exposure times are summarised in \cref{tab:Lya_fluxes}.

Finally, we revisit the medium-resolution NIRSpec G235M/F170LP observations from PID 1287, which in \citetalias{2025Natur.639..897W} only comprised $4.7 \, \mathrm{h}$ due to the same guide-star lock failure, but are now completed to $6.9 \, \mathrm{h}$ depth as initially scheduled. Furthermore, they are complemented with another $9.3 \, \mathrm{h}$ of G235M data from a pilot NIRSpec dense-shutter spectroscopy (DSS) survey, obtained as part of rescheduled observations from the JOF programme \citep{2026MNRAS.549ag824D}. However, due to the $30\%$ reduced line-flux sensitivity of the DSS observing strategy\footnote{This is a result of the marginally higher background level, which is counteracted in terms of survey speed by a $4$--$5\times$ increase in target density \citep[cf.][]{2026MNRAS.549ag824D}.}, the effective exposure time for the DSS observations is $0.7^2 \times 9.3 \, \mathrm{h} \approx 4.6 \, \mathrm{h}$. Altogether, this results in an expected $\sqrt{(6.9 \, \mathrm{h} + 4.6 \, \mathrm{h})/4.7 \, \mathrm{h}} \approx 1.56\times$ increase in sensitivity with respect to \citetalias{2025Natur.639..897W}.

\subsection{JWST/NIRSpec data reduction}
\label{ssec:Data_reduction}

For all available NIRSpec spectroscopy, we used the NIRSpec GTO Collaboration data reduction pipeline \texttt{v5.1}, as described in DR4 paper II \citep{2026MNRAS.549ag939S}. Compared to the previous \texttt{v3} pipeline adopted in DR3 \citep{2025ApJS..277....4D} and the results presented in \citetalias{2025Natur.639..897W}, the main changes are to adopt more recent NIRSpec calibration files and to include improved wavelength and flux calibration corrections depending on intra-shutter offsets. Given the spatial compactness of \JGSzthirteenLA \citep{2024ApJ...970...31R}, we default to one-dimensional spectra extracted from a 3-pixel aperture to optimise the SNR. As in the Supplementary information of \citetalias{2025Natur.639..897W}, however, we have verified that our findings on the spectral shape and normalisation are robust against changing to the standard full (5-pixel) aperture. This also holds for the line emission, including in a reduction where only the two outer nod positions are considered (instead of the default three, which may cause self-subtraction if the emission extends beyond the central micro-shutter; \cref{app:Grating_spectra}). For the PRISM spectroscopy, a total of $144$ sub-exposures from PIDs 1287 and 5997 ($72$ each) are available with a total exposure time of $56 \, \mathrm{h}$ (\cref{tab:Lya_fluxes}). These one-dimensional sub-exposures were filtered and combined using the prescription described in \citet{2024ApJ...976..160H}, \citetalias{2025Natur.639..897W}, and \citet{2026OJAp....955261W}, as part of which we also constructed the covariance matrix in a bootstrapping procedure with $5000$ iterations.

\begin{figure*}
	\centering
	\includegraphics[width=\linewidth]{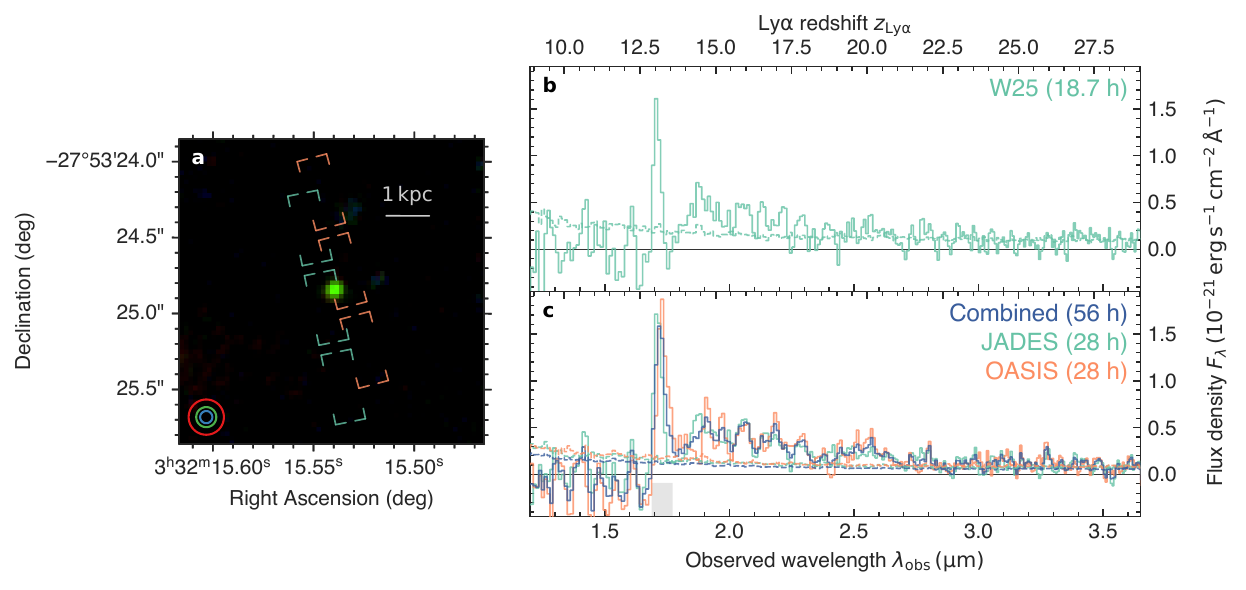}
	\caption{NIRCam and NIRSpec/PRISM observations of \JGSzthirteenLA. (a) False-colour NIRCam image with the F090W, F115W, F150W, and F162M filters shown as blue, F182M, F200W, F210M, F250M, and F277W as green, and F300M, F335M, F356W, F410M, F444W, and F480M as red (stacked at native resolution). For reference, the point spread function FWHM of the F115W, F200W, and F356W filters are shown in the bottom-left corner. The triplet of NIRSpec micro-shutters that were opened in the JADES and OASIS observations are shown in teal and orange, respectively. A physical scale of $1 \, \mathrm{kpc}$ at $z = 13$ is indicated by the horizontal white line. (b) One-dimensional NIRSpec/PRISM spectrum from JADES \citepalias{2025Natur.639..897W}. (c) PRISM spectra from the completed JADES (teal) and new OASIS (orange) observations, and the combined spectrum (dark blue). The grey shading indicates the wavelength range over which the \Lya line fluxes are integrated (\cref{tab:Lya_fluxes}). $1 \sigma$ uncertainty on individual wavelength bins is shown by dashed lines in panels~b and c.
	}
	\label{fig:Spectra_comparison}
\end{figure*}

We performed this combination process with all $144$ PRISM sub-exposures to obtain a final, one-dimensional spectrum. Using the same procedure, we additionally created combined spectra after having split up the dataset into the $72$ sub-exposures taken as part of PID 1287 (JADES) and 5997 (OASIS). A comparison of these separate JADES and OASIS spectra will be discussed further in \cref{ssec:Comparison_DEEP-JWST_OASIS}, especially in light of the different intra-shutter positions between these two observations, both near the edge of one of NIRSpec's micro-shutters, but critically at almost opposite ends along the dispersion direction (\cref{fig:Spectra_comparison}a).

\section{Results}
\label{sec:Results}

\subsection{Comparison of JADES and OASIS observations}
\label{ssec:Comparison_DEEP-JWST_OASIS}

\subsubsection{A first look at the continuum properties}
\label{sssec:Comparison_DEEP-JWST_OASIS_continuum}

The resulting one-dimensional NIRSpec/PRISM spectra from the JADES and OASIS observations are shown in \cref{fig:Spectra_comparison}b and \cref{fig:Spectra_comparison}c. From this comparison it becomes clear that, especially considering the estimated uncertainties, the new OASIS data show excellent agreement with the spectrum based on (the initial $67\%$ of) the JADES observations previously reported in \citetalias{2025Natur.639..897W}. Both the now complete JADES spectrum and the new OASIS spectrum confirm the overall continuum shape with a clear spectral break at $\lambda_\text{obs} \approx 1.7 \, \mathrm{\upmu m}$, as well as its normalisation.

More specifically, we have verified that synthetic NIRSpec photometry in both the OASIS and JADES spectroscopy continues to agree well with NIRCam photometry. Indeed, with the updated NIRSpec calibration files (\cref{ssec:Data_reduction}) the agreement is slightly improved compared to what was found in \citetalias{2025Natur.639..897W}, though we still find the NIRSpec synthetic photometry to be systematically lower, by approximately $20\%$. As before, this small residual difference does not display a clear trend with wavelength, and this is within the level of systematic uncertainty expected from the finite pointing accuracy and substantial gradient in slit losses near the edge of the micro-shutter\footnote{The MSA pointing accuracy is $\ssim 20 \, \mathrm{mas}$, which at the edge of the micro-shutter translates into transmission differences of $\gtrsim 10\%$ \citep{2022A&A...661A..81F}.}, where \JGSzthirteenLA is observed (see \cref{fig:Spectra_comparison}a). Unless mentioned otherwise, we therefore apply a minor additional path-loss correction to both spectra in the following, updated to the new spectroscopic measurements but otherwise using the same routine as described in \citetalias{2025Natur.639..897W}. We use the NIRCam photometry obtained by \textsc{forcepho} \citep[B.~D.~Johnson et al. in prep.; see also e.g.][]{2023NatAs...7..611R, 2023ApJ...952...74T, 2025NatAs...9..141B}, noting that the regular aperture photometry yields consistent results.
\begingroup
    \setlength{\tabcolsep}{6pt}
    \renewcommand{\arraystretch}{1.25}
    \begin{table}[t]
        \centering
        \footnotesize
        \caption{Summary of NIRSpec observations of \JGSzthirteenLA as part of the JADES, JOF, and OASIS programmes (\cref{ssec:JWST_observing_programmes}).}
        \begin{tabular}{lllll}
            \toprule
             & JADES & OASIS & JOF & Combined \\
            \midrule
            PID & 1287 & 5997 & 3215 & -- \\
            $t_\text{exp, PRISM} \, (\mathrm{h})$ & $28$ & $28$ & -- & $56$ \\
            $t_\text{exp, G235M} \, (\mathrm{h})$ & $6.9$ & -- & $9.3$\tablefootmark{$\ast$} & $16.2$\tablefootmark{$\ast$} \\
            $F_{19}$ & $6.3 \pm 1.0$ & $7.9 \pm 1.2$ & -- & $7.1 \pm 0.8$ \\
            $\text{SNR}$ & $6.1$ & $6.4$ & -- & $8.8$ \\
            \bottomrule
        \end{tabular}
        \tablefoot{
            Listed properties are programme ID (PID), exposure time ($t_\text{exp}$) in the PRISM and G235M, and simple estimates from the PRISM spectra of the line flux in units of $10^{-19} \, \mathrm{erg \, s^{-1} \, cm^{-2}}$ ($F_{19}$) and SNR. \\
            \tablefoottext{$\ast$}{With (partially) increased background noise level (see text for details).}
        }
        \label{tab:Lya_fluxes}
    \end{table}
\endgroup

\subsubsection{Initial \texorpdfstring{\Lya}{\Lyatext} line measurements}
\label{sssec:Comparison_DEEP-JWST_OASIS_continuum_line}

Crucially, the NIRSpec/PRISM observations from OASIS also reaffirm the bright emission line. More quantitatively, we confirm that the emission line fluxes seen in the JADES and OASIS data agree within uncertainties, as summarised in \cref{tab:Lya_fluxes}. We obtained line fluxes by direct integration after path-loss correction and continuum subtraction\footnote{Based on the best-fit continuum obtained from the full spectral modelling in \cref{ssec:Spectral_modelling_setup,ssec:Spectral_modelling_results}.}, having selected $7$ wavelength bins where the combined spectrum reaches $\text{SNR} > 3$ for ease of comparison (grey shading in \cref{fig:Spectra_comparison}c; more sophisticated estimates will be provided in \cref{ssec:Spectral_modelling_results}). Moreover, we find that the integrated difference between the two spectra across the same wavelength bins is statistically consistent with zero, $\Delta F = (1.35 \pm 1.32) \times 10^{-19} \, \mathrm{erg \, s^{-1} \, cm^{-2}}$.

While the line was already significantly detected by \citetalias{2025Natur.639..897W} in the initial $18.7 \, \mathrm{h}$, a particularly encouraging trend emerges from the integrated line SNRs. The initial \citetalias{2025Natur.639..897W} dataset shows $\text{SNR} \approx 5$ (re-measured here with a consistent continuum-subtraction method), which is seen to increase from $\text{SNR} \approx 6$ in both the complete JADES and OASIS spectra ($28 \, \mathrm{h}$ each), and finally $\text{SNR} \approx 9$ for the combined $56 \, \mathrm{h}$ of observations. The scaling is in lockstep with the square root of the exposure time, precisely as expected for a real signal.

We recall that, as discussed in \citetalias{2025Natur.639..897W}, the observed discontinuity at $1.7 \, \mathrm{\upmu m}$---more than $3 \, \mathrm{mag}$ between the NIRCam F150W and F200W filters---leaves no other credible alternative than for the redshift to be $z \approx 13$, and hence for the line to be \Lya. Furthermore, line emission from a low-redshift interloper (e.g. \Halpha at $z \approx 1.60$, $\OIII \, \lambda \, 5008 \, \Angstrom$ at $z \approx 2.41$, or $\OII \, \lambda \, 3727, 3730 \, \Angstrom$ at $z \approx 3.58$) would require both an unprecedented EW (observed $\text{EW} \gtrsim \num{10000} \, \Angstrom$) as well as unphysical line ratios to explain the absence of other strong lines (e.g. \Paalpha, \Hbeta, and \OIII) that are expected to be present for all of these redshift solutions.

Having confirmed the line in the new PRISM observations, we considered the deeper NIRSpec/G235M data, consisting of a $9.3 \, \mathrm{h}$ exposure from the DSS survey presented in \citet{2026MNRAS.549ag824D} and the now fully completed $6.9 \, \mathrm{h}$ obtained as part of JADES (\cref{ssec:JWST_observing_programmes}). Encouragingly, compared to medium-resolution G235M observations considered in \citetalias{2025Natur.639..897W}, a small flux excess does indeed appear within a spectral aperture of $\Delta v \approx \num{1000} \, \mathrm{km \, s^{-1}}$ centred on $\lambda_\text{obs} = 1.7171 \, \mathrm{\upmu m}$ (the central wavelength inferred from the PRISM spectra; \cref{ssec:Spectral_modelling_results}). While the excess is still decidedly marginal (just over $1\sigma$), we have verified that this is in line with expectations based on the PRISM flux measurements, if the line is spectrally resolved in the $R \approx 1000$ observations (\cref{app:Grating_spectra}).

An upcoming Cycle-4 programme (PID 9016, PI: D.~Stark) will target \JGSzthirteenLA, among other notable galaxies around GOODS-S, using G235M/F170LP for an additional $45 \, \mathrm{h}$, nearly quintupling the total exposure time. Provided the intrinsic line full width at half maximum is not too large ($\text{FWHM} \lesssim \num{1000} \, \mathrm{km \, s^{-1}}$), our current sensitivity levels (\cref{app:Grating_spectra}) suggest that this should provide sufficient depth for a more robust detection of the line and shed light on its width.

Closer examination of the spectral shape of the line observed in the PRISM observations reveals that its peak appears ever so slightly offset between the two separate spectra, of the order of a single wavelength bin. We deem it unlikely to be a real physical effect due to intrinsic time variability, given the relatively short delay in the rest frame of approximately one month between the JADES and OASIS observations. Another physical explanation is that the \Lya emission, contrary to the UV continuum, could be spatially extended \citep[as observed in e.g. GN-z11 at $z = 10.6$;][]{2023A&A...677A..88B, 2024A&A...687A.283S}, possibly causing the two different micro-shutter placements to trace distinct sub-regions where the line exhibits different peak velocity offsets \citep[e.g.][]{2020A&A...635A..82L}. However, we do not see evidence for spatially extended emission in the two-dimensional spectra (\cref{app:Grating_spectra}), and the implied velocity difference still appears unrealistically large, since the observed wavelength of the line coincides with the lowest spectral resolution of the PRISM (where individual bins have a velocity width of $\Delta v \approx \num{2000} \, \mathrm{km \, s^{-1}}$; \citealt{2022A&A...661A..80J}). We instead attribute this to a combination of relatively modest SNR and the changing intra-shutter positions of \JGSzthirteenLA, which were nearly at opposite ends of the NIRSpec micro-shutter between the two programmes, for reasons discussed in more detail below.

Wavelength offsets in NIRSpec/PRISM at the level of $\Delta v \sim \num{1000} \, \mathrm{km \, s^{-1}}$ have been reported previously in sources where accurate redshift measurements are available from the higher-resolution gratings \citep[$R \approx \num{1000}$; e.g.][]{2024A&A...690A.288B, 2025A&A...697A.189D}. Based on JADES DR3 data, \citet{2025ApJS..277....4D} demonstrated that an exceptionally clear correlation ($p < 10^{-30}$) exists between these offsets and the intra-shutter position of sources along the dispersion direction. As such, the effect can be interpreted as small imperfections of the NIRSpec wavelength solution for sources that are offset from the micro-shutter centre. While the standard NIRSpec data reduction algorithms\footnote{These algorithms are shared between the standard Space Telescope Science Institute (STScI; \url{https://jwst-docs.stsci.edu/jwst-science-calibration-pipeline}) and the NIRSpec GTO data reduction pipelines (\cref{ssec:Data_reduction} and \citealt{2026MNRAS.549ag939S}).} already include a zero-point wavelength refinement as a function of intra-shutter source displacement, until recently this calibration was entirely based on pre-flight instrument models \citep{2018SPIE10704E..0QA, 2022A&A...661A..81F}.

Steadily growing datasets, however, have prompted efforts to empirically calibrate any remaining inaccuracies, including within JADES DR4 \citep{2026MNRAS.549ag939S}. However, the blue wavelength range of NIRSpec ($\lambda_\text{obs} \lesssim 2 \, \mathrm{\upmu m}$) poses the most difficult regime to perform this correction, since this requires a large number of emission lines detected at high SNR, ideally from sources that remain (close to being) spatially unresolved. Although strong optical lines (e.g. $\OIII \, \lambda \, 4960, 5008 \, \Angstrom$, \Hbeta, \Halpha) can be readily observed in galaxies at $z \lesssim 3$, the JADES target selection was designed to favour higher-redshift sources \citep{2024A&A...690A.288B, 2026MNRAS.549ag836C}, resulting in reduced number statistics below $\ssim 2 \, \mathrm{\upmu m}$ available for this analysis \citep[Fig.~5 in][]{2026MNRAS.549ag939S}. Furthermore, these lower-redshift sources are typically spatially extended, thus likely hindering an accurate empirical determination of the wavelength zero-point calibration for a given intra-shutter position.

Statistically speaking, the empirically calibrated routine implemented within the \texttt{v5.1} NIRSpec GTO pipeline yields a clear improvement to the PRISM wavelength offsets, as shown in Fig.~4 of \citet{2026MNRAS.549ag939S}. Due to the limited statistics at $\lambda_\text{obs} \lesssim 2 \, \mathrm{\upmu m}$, however, minor discrepancies of the order of $\ssim \num{1000} \, \mathrm{km \, s^{-1}}$ remain in individual sources, as is also the case in the peak \Lya wavelength between the OASIS and JADES observations (\cref{fig:Spectra_comparison}). Recalling the absence of significant differences in the \Lya emission between different spectral extraction apertures (\cref{ssec:Data_reduction}), which suggests that the line emission is likely spatially compact, we conclude that remaining calibration systematics could reasonably cause the observed wavelength shift. In the following section, we therefore turn to the Bayesian spectral fitting method presented in \citetalias{2025Natur.639..897W} to marginalise over the remaining wavelength discrepancy between the JADES and OASIS observations.
\begin{figure*}
	\centering
    \sidecaption
	\includegraphics[width=12cm]{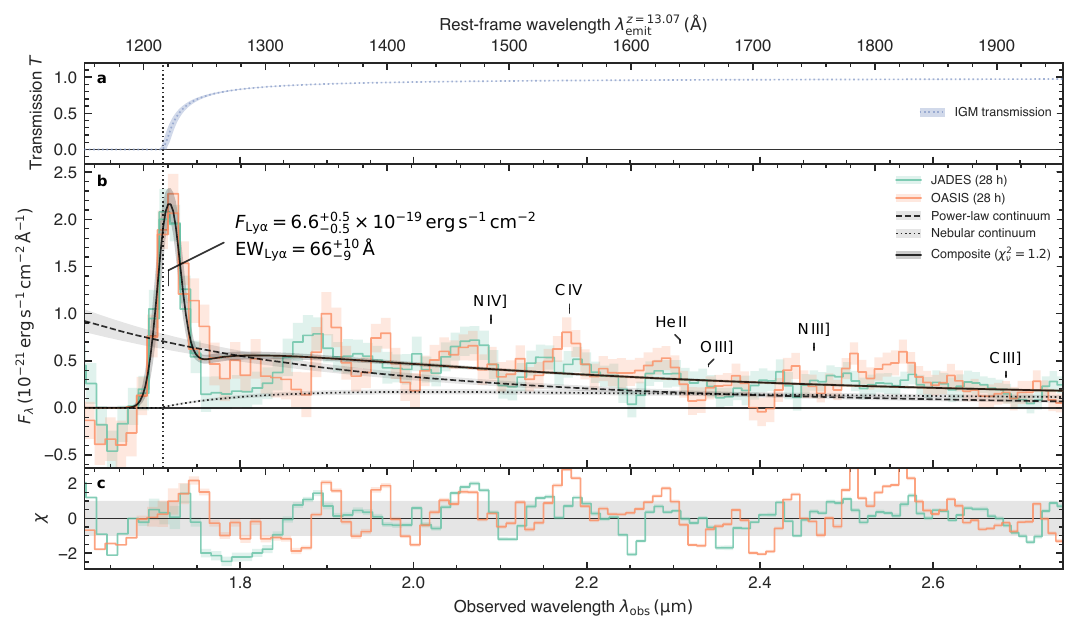}
	\caption{Modelled NIRSpec/PRISM observations of \JGSzthirteenLA. (a) Model curves for the IGM transmission $T$, according to the legend on the right. (b) Coloured lines show the path-loss corrected PRISM spectra observed in JADES and OASIS (\cref{ssec:JWST_observing_programmes}). The best-fit model spectrum is shown by the black solid line, while its individual power-law and nebular-continuum components are shown with dashed and dotted lines, respectively. The observed \Lya flux and EW in the best-fit model are annotated. The expected location of rest-frame UV lines is indicated for the model's best-fit systemic redshift of $z = 13.07$, although we note that none are confidently detected (\cref{sec:Discussion}). (c) Uncertainty-normalised residuals $\chi$ are shown for the two different spectra.
	}
	\label{fig:Model_spectra}
\end{figure*}

\subsection{Spectral line and continuum modelling: model set-up}
\label{ssec:Spectral_modelling_setup}

To reassess the line and continuum properties of \JGSzthirteenLA from the full $56 \, \mathrm{h}$ of observations, we repeated the spectral modelling presented in \citetalias{2025Natur.639..897W}.\footnote{Code available at \url{https://github.com/joriswitstok/lymana_absorption}.} We refer to the Methods section of that work for further details, before briefly summarising the model set-up and outlining minor changes below. When comparing model variants to each other, we will explicitly define a shorthand for each variant.
\begingroup
    \setlength{\tabcolsep}{3.75pt}
    \renewcommand{\arraystretch}{1.5}
    \begin{table}[t]
        \centering
        \footnotesize
        \caption{Spectral line and continuum modelling results.}
        \begin{tabular}{lll}
            \toprule
            Parameter & (Logarithmic) units & Value
            \\
            \midrule
            $z_\text{sys}$ & & $13.07_{-0.02}^{+0.04}$
            \\
            $C$ & $10^{-21} \, \mathrm{erg \, s^{-1} \, cm^{-2} \, \Angstrom^{-1}}$ & $0.25_{-0.03}^{+0.03}$
            \\
            $\beta_\text{UV}$ & & $< -3.27$ ($3\sigma$)
            \\
            $R_\text{ion}$ & $\mathrm{pMpc}$ & $0.167_{-0.062}^{+0.070}$
            \\
            $\log_{10} (\xi_\text{ion})$ & $\mathrm{Hz \, erg^{-1}}$ & $26.43_{-0.15}^{+0.30}$
            \\
            $f_\text{esc, LyC}$ & & $0.55_{-0.29}^{+0.25}$
            \\
            $F_\text{\Lya}$ & $10^{-19} \, \mathrm{erg \, s^{-1} \, cm^{-2}}$ & $6.6_{-0.5}^{+0.5}$
            \\
            $L_\text{\Lya}$ & $10^{43} \, \mathrm{erg \, s^{-1}}$ & $1.1_{-0.2}^{+0.2}$
            \\
            $\text{EW}_\text{\Lya, pre-IGM}$ & $\Angstrom$ & $431_{-84}^{+106}$
            \\
            $\text{EW}_\text{\Lya}$ & $\Angstrom$ & $66_{-9}^{+10}$
            \\
            $f_\text{esc, \Lya}$ & & $0.152_{-0.020}^{+0.024}$
            \\
            $\Delta v_\text{\Lya, pre-IGM}$ & $\mathrm{km \, s^{-1}}$ & $843_{-174}^{+107}$
            \\
            $\Delta v_\text{\Lya}$ & $\mathrm{km \, s^{-1}}$ & $1164_{-658}^{+686}$
            \\
            $\sigma_\text{\Lya, pre-IGM}$ & $\mathrm{km \, s^{-1}}$ & $448_{-85}^{+77}$
            \\
            $\Delta p_0$ & $\mathrm{bins}$ & $0.55_{-0.29}^{+0.36}$
            \\
            $\Delta p_1$ & $\mathrm{bins}$ & $-0.62_{-0.33}^{+0.33}$
            \\
            \midrule
            $\chi_\nu^2$ & & $1.22$
            \\
            BIC & & $350$
            \\
            \bottomrule
        \end{tabular}
        \tablefoot{
            Reported model parameters are the systemic redshift ($z_\text{sys}$), normalisation ($C$) and slope ($\beta_\text{UV}$) of the power-law continuum, LyC production efficiency ($\xi_\text{ion}$) and escape fraction ($f_\text{esc, LyC}$), the \Lya peak velocity offset ($\Delta v_\text{\Lya, pre-IGM}$) and velocity dispersion ($\sigma_\text{\Lya, pre-IGM}$) before IGM transmission (i.e. as the line profile emerges from the galaxy), and the two wavelength corrections $\Delta p_i$. Additionally, we report the $\chi_\nu^2$ and BIC values, as well as the following quantities not freely varied but derived from the main parameters \citepalias[see \cref{ssec:Spectral_modelling_setup} and][]{2025Natur.639..897W}: the ionised bubble radius ($R_\text{ion}$), observed \Lya flux ($F_\text{\Lya}$), intrinsic \Lya luminosity ($L_\text{\Lya}$), the EW of \Lya before IGM transmission ($\text{EW}_\text{\Lya, pre-IGM}$) and afterwards (as observed; $\text{EW}_\text{\Lya}$), the escape fraction of \Lya ($f_\text{esc, \Lya}$), and the observed \Lya velocity offset ($\Delta v_\text{\Lya}$).
        }
        \label{tab:Model_results}
    \end{table}
\endgroup

The fitting routine aims to reproduce the observed \Lya and continuum emission within the observed wavelength range $1.6 \, \mathrm{\upmu m}$  to $2.9 \, \mathrm{\upmu m}$ (where the continuum is detected at $\text{SNR} \gtrsim 1$ per wavelength bin). The model takes into account instrumental effects in the form of the NIRSpec/PRISM line spread function\footnote{Based on the \href{https://jwst-docs.stsci.edu/jwst-near-infrared-spectrograph/nirspec-instrumentation/nirspec-dispersers-and-filters}{resolution curve in the STScI documentation}, enhanced by $1.5\times$ to match the \Lya line width in the PRISM as described in \citetalias{2025Natur.639..897W}.}, while physical effects include the \Lya damping-wing absorption from intervening neutral hydrogen in the IGM, which is assumed to be neutral except for a potential ionised bubble, as well as from potential proximate DLA systems. Ionising photons, whose rate is parametrised by the production efficiency $\xi_\text{ion}$, contribute to \Lya visibility either through (i) direct conversion into \Lya photons via recombinations, or (ii) indirect \Lya transmission enhancement via ionised bubble growth. The relative importance of these two processes is modulated by the LyC escape fraction of the galaxy, $f_\text{esc, LyC}$ (in the default single-source model). In addition, we considered a more agnostic approach here where we simply vary the ionised bubble radius with a log-uniform prior, $\log_{10} R_\text{ion} \, (\mathrm{pMpc}) \sim \mathcal{U}(-2, 0)$ \citep[the upper bound of $1 \, \mathrm{pMpc}$ being motivated by simulation predictions at a neutral fraction of $x_\text{\HI} \gtrsim 80\%$;][]{2024MNRAS.528.4872L, 2024MNRAS.531.2943N}, without requiring any LyC leakage from \JGSzthirteenLA itself (collective model). Finally, as in \citetalias{2025Natur.639..897W} we also considered a variant of this model without an ionised bubble (i.e. fixing $R_\text{ion} = 0$ as well as $f_\text{esc, LyC} = 0$; no-bubble model).

We adopted the same (log-)uniform prior distributions on our fitting parameters as in \citetalias{2025Natur.639..897W} except for the systemic redshift, here slightly expanded to $z_\text{sys} \sim \mathcal{U}(12.7, 13.5)$. In addition, the prior on the \Lya velocity offset (before encountering the neutral IGM) was previously assumed to follow $\Delta v_\text{\Lya, pre-IGM} \sim \mathcal{U}(0, 500) \, \mathrm{km \, s^{-1}}$, but we conservatively extend its upper bound here to $\num{1000} \, \mathrm{km \, s^{-1}}$. Based on the absolute UV magnitude of \JGSzthirteenLA \citepalias[$M_\text{UV} \approx -18.5 \, \mathrm{mag}$;][]{2025Natur.639..897W}, such a high \Lya velocity offset would be unlikely, but perhaps not entirely unfeasible, when compared to observations ranging from $z \sim 2$ \citep{2014ApJ...795...33E, 2014ApJ...795..165S} to $z \sim 11$ (see e.g. \citealt{2018ApJ...856....2M}; \citealt{2025MNRAS.538L..16Q} for literature compilations). Observational evidence has moreover long suggested that, especially at $z \gtrsim 5$, \Lya line profiles could have prominent non-Gaussian scattering wings, which may extend out to several hundreds of $\mathrm{km \, s^{-1}}$ redwards of line centre \citep{2003MNRAS.342L..47B, 2014ApJ...795...33E, 2017MNRAS.467.3306S, 2018A&A...619A.147P, 2021ApJ...908...36H, 2021MNRAS.508.1686W, 2024MNRAS.531.2701T, 2025MNRAS.542..762Y, 2026A&A...712A.113P}. The main effect of this extended prior range is a larger uncertainty on the inferred systemic redshift, which remains unclear despite the $56 \, \mathrm{h}$ depth of the complete JADES and OASIS observations, as discussed further below.

Following \citet{2025A&A...702A..57H}, as a means of refined wavelength correction we applied a constant shift of $\Delta p_i$ in units of wavelength bins, treating the JADES ($i = 0$) and OASIS ($i = 1$) spectra separately. This results in an absolute wavelength correction most notable around $\lambda_\text{obs} \approx 1.6 \, \mathrm{\upmu m}$, where the spectral resolution of the NIRSpec/PRISM is lowest.\footnote{The reason for working in wavelength bins (whose size is tied to $1.59\times$ the width of a native detector pixel in the NIRSpec GTO pipeline) is the wavelength-dependent spectral resolution of the PRISM, which means that they sample an irregular grid in wavelength space \citep[; see also e.g. \citealt{2025A&A...697A.189D}; \citealt{2026MNRAS.549ag939S}]{2022A&A...661A..80J}.} We adopted a Gaussian prior with $\mu_{\Delta p_i} = 0$ and $\sigma_{\Delta p_i} = 0.5$, truncated at $|\Delta p_i| = 2$.

\subsection{Spectral modelling: insights and results}
\label{ssec:Spectral_modelling_results}

Firstly, we recall that the aim of the spectral modelling approach adopted here is not to provide the most physically detailed model possible. While full hydrodynamic radiative-transfer simulations are able to capture many of the relevant physical processes governing the production and escape of \Lya photons \citep[e.g.][]{2009ApJ...704.1640L, 2019MNRAS.484...39S, 2023MNRAS.523.3749B}, they also require a variety of underlying assumptions that shape the distributions of gas and dust in galaxies. Conversely, the current phenomenological model is designed to obtain purely empirical constraints on the basic physical properties that \JGSzthirteenLA must possess to explain the observed strong \Lya emission. In the following, we discuss various aspects of the model and how they compare to the findings presented in \citetalias{2025Natur.639..897W} previously.

\subsubsection{Continuum emission}
\label{sssec:Spectral_modelling_results_continuum}

As in \citetalias{2025Natur.639..897W}, we considered various combinations of a power law and/or a nebular continuum as the underlying continuum emission: a pure power-law continuum, a pure two-photon continuum, and a composite model with a power-law and nebular continuum. To reproduce the continuum downturn near \Lya, we explored models where the power-law continuum is affected by DLA absorption (IGM and DLA model) in addition to the fiducial pure IGM absorption model. However, noting that the turnover near \Lya is less pronounced in the new OASIS spectrum compared to JADES---that is, it does not tend to zero at the \Lya wavelength to the same degree---we tested the introduction of a covering fraction of the DLA-absorbing gas in the IGM and DLA model. The associated geometry, where a central source is only partially covered by dense neutral \HI gas (e.g. an edge-on disc), would also provide a natural explanation for the simultaneous, efficient escape of \Lya photons, as discussed in \citetalias{2025Natur.639..897W} (see also \cref{ssec:Ionising_source_and_ISM_conditions_within_GS-z13-1}).

Statistically, however, based on the combined JADES+OASIS data we find that a pure IGM absorption model produces as satisfactory a fit as a model with DLA absorption (IGM and DLA model; both a reduced chi-squared value of $\chi_\nu^2 = 1.22$). In terms of the Bayesian information criterion (BIC), the former is slightly preferred ($\text{BIC} = 350$ versus $\text{BIC} = 360$, respectively), although we note that this is mainly true with the composite model, whereas in the pure power-law model the preference becomes less strong. For simplicity, we therefore consider the family of models without DLA absorption in what follows, although we note that it is difficult to conclusively rule out the presence of DLA absorption altogether (as discussed further in \cref{ssec:Ionising_source_and_ISM_conditions_within_GS-z13-1}).

On the other hand, we confirm that a pure two-photon continuum continues to be disfavoured over the pure power-law model ($\chi_\nu^2 = 1.37$ versus $\chi_\nu^2 = 1.22$ and $\text{BIC} = 382$ versus $\text{BIC} = 349$, respectively), owing to the very blue continuum of \JGSzthirteenLA. Otherwise, however, we see little difference statistically between power-law continuum models either with the addition of a nebular continuum (composite model), or without it (pure power-law model; both $\chi_\nu^2 = 1.22$, $|\Delta \text{BIC}| < 1$). As our fiducial model, we opt here for the case where nebular continuum is included. The best-fitting results for this $9$-parameter model are shown in \cref{fig:Model_spectra} and summarised in \cref{tab:Model_results}. The full posterior distributions are presented in \cref{app:Corner_plot}.

We recall that the normalisation of the nebular continuum is self-consistently coupled to the \Lya luminosity, which itself is controlled by $\xi_\text{ion}$ (and $f_\text{esc, LyC}$ in the single-source ionised bubble model). We computed nebular continuum templates with the \textsc{pyneb} code \citep{2015A&A...573A..42L} under the assumption of $T = \num{20000} \, \mathrm{K}$ and $n = 100 \, \mathrm{cm^{-3}}$, noting that the shape of the $2\gamma$ component---which dominates at the wavelength regime considered here---is fixed \citep{1951ApJ...114..407S}. The inclusion of the nebular continuum effectively acts to redden the power-law continuum, which motivates the choice for this relatively low density here as effectively the bluest UV slope, though this will be explored further in \cref{ssec:Ionising_source_and_ISM_conditions_within_GS-z13-1}. As a consequence, even though the effective UV slope (power-law and nebular continuum combined) is $\beta_\text{UV} = -3.03$, the intrinsic UV slope is preferred to be extremely steep, $\beta_\text{UV} < -3.27$ ($3\sigma$). The pure power-law model, meanwhile, yields a very blue value still, $\beta_\mathrm{{UV}} = -3.21 \pm 0.28$. We have verified that this measurement is consistent within uncertainties compared to variations of fitting the 3-pixel versus 5-pixel spectral extraction (\cref{ssec:Data_reduction}), (not) including DLA absorption, and/or extending the wavelength range considered.

\subsubsection{Ionised bubble}
\label{sssec:Spectral_modelling_results_bubble}

In \citetalias{2025Natur.639..897W}, it was concluded that the scenario without an ionised bubble is strongly disfavoured by the non-detection in the MIRI/F770W band containing \Hbeta. Even under conservative assumptions (no continuum or $\OIII \, \lambda \, 4960, 5008 \, \Angstrom$ emission contaminating the F770W filter), this translates to a $3\sigma$ upper bound on the \Hbeta flux of $F_\text{\Hbeta} \lesssim 6.7 \times 10^{-19} \, \mathrm{erg \, s^{-1} \, cm^{-2}}$, and hence a \Lya luminosity of $L_\text{\Lya} \lesssim 3.7 \times 10^{43} \, \mathrm{erg \, s^{-1}}$ (case-B recombination). Here, we find that the no-bubble model results in a \Lya luminosity of $(1.4 \pm 0.1) \times 10^{43} \, \mathrm{erg \, s^{-1}}$, nominally consistent with the MIRI/F770W non-detection. That is, when we correct the observed \Lya strength to its intrinsic value accounting for the modelled IGM transmission, and subsequently convert it to an \Hbeta flux under case-B recombination, it would produce an F770W signal at approximately the $1.1\sigma$ level (i.e. not significantly detectable).

However, the resulting \Lya velocity offset---which is degenerate with, and therefore the main factor controlling, the inferred ionised bubble size, as discussed below---is found to be pushed to the very edge of the prior range, $\Delta v_\text{\Lya, pre-IGM} = 959_{-50}^{+29} \, \mathrm{{km \, s^{{-1}}}}$. This explains how the formal F770W tension is reduced compared to \citetalias{2025Natur.639..897W}, since the prior on the velocity offset was more conservatively extended redwards here (\cref{ssec:Spectral_modelling_setup}), and it implies that the complete absence of an ionised bubble would require a highly fine-tuned scenario: in addition to the \Hbeta flux dominating the MIRI/F770W filter, the \Lya line profile would need to be extremely skewed towards the red. Additionally, if half of \Lya emerges closer to systemic where it is suppressed by the IGM, or if the continuum and \OIII lines contribute equally to the observed F770W flux---either of which would double the implied F770W flux---we would have expected a $\ssim2\sigma$ F770W signal. We therefore conclude that the presence of an ionised bubble is difficult to prove unambiguously, but we deem it the more likely explanation for the observed luminous \Lya line.

In the default model, on the other hand, we can deduce from the dependencies of the posterior distributions shown in \cref{fig:Corner_plot} that $R_\text{ion}$ is not strongly sensitive to the line width or the wavelength corrections. However, $R_\text{ion}$ is anti-correlated with the velocity offset, as expected from the damping-wing absorption curve rising monotonically redwards of systemic (\cref{fig:Model_spectra}a). It should be kept in mind, therefore, that the inference on ionised bubble sizes is model-dependent and should be treated as suggestive. In the default model, the velocity offset is found to be bounded from below with a minimum of $\Delta v_\text{\Lya, pre-IGM} \approx 400 \, \mathrm{km \, s^{-1}}$, a constraint which is mainly tied to the fact that \JGSzthirteenLA is assumed to create its own ionised bubble in the default model (a lower velocity offset is allowed in the collective model, discussed below). The weak or absent constraints on the intrinsic line width, meanwhile, are likely an effect of the IGM damping-wing absorption, which causes any intrinsic line shape to be effectively broadened (see e.g. Appendix A in \citealt{2024A&A...682A..40W}).

The collective model, which does not require any LyC leakage from \JGSzthirteenLA but instead simply imposes an ionised bubble to be in place, yields virtually identical results both in terms of fit quality and (most) best-fit parameters as those of the single-source model. The two exceptions are the velocity offset, which is no longer pushed against the upper edge of the prior range ($\Delta v_\text{\Lya, pre-IGM} = 703_{-353}^{+213} \, \mathrm{km \, s^{-1}}$, i.e. $\lesssim 400 \, \mathrm{km \, s^{-1}}$ within $1\sigma$ uncertainties), and the ionising-photon production efficiency, which is reduced by $0.5 \, \mathrm{dex}$ to $\log_{{10}} \xi_\mathrm{{ion}} \, (\mathrm{{Hz \, erg^{{-1}}}}) = 25.91_{-0.22}^{+0.12}$, although both of these come at the expense of requiring a larger ionised bubble, $R_\mathrm{{ion}} = 0.48_{-0.33}^{+0.24} \, \mathrm{{pMpc}}$. Therefore, the collective model carries the implicit assumption that the environment of \JGSzthirteenLA has managed to ionise a significant portion of the surrounding IGM. Whether or not this is a realistic scenario will be further explored in \cref{ssec:Environment}.
\begin{figure*}
	\centering
	\includegraphics[width=\linewidth]{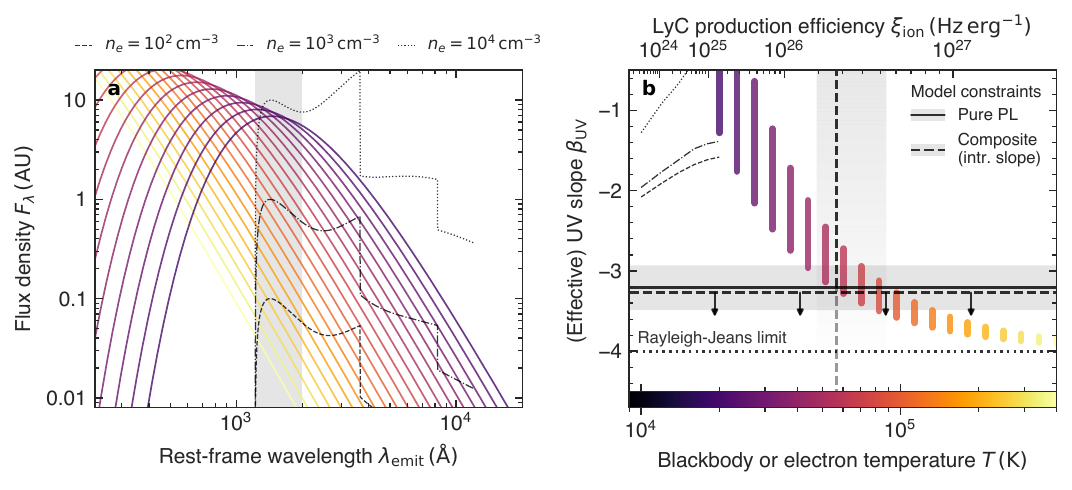}
	\caption{(a) Coloured lines show blackbody curves of varying temperature (according to the colour bar in panel~b). Thin dashed, dash-dotted, and dotted lines show nebular continuum predictions by \textsc{pyneb} \citep{2015A&A...573A..42L} for varying electron density $n_e$, but at fixed temperature $T_e = \num{10000} \, \mathrm{K}$ (where they are bluest; cf. panel~b). Grey shading indicates the approximate wavelength range probed in the spectral modelling of \JGSzthirteenLA (\cref{ssec:Spectral_modelling_setup}). (b) Effective UV slopes of blackbody spectra and nebular continua measured across the wavelength range highlighted in panel~a, as a function of temperature. Thin lines show the minimum slope of nebular continua at different densities (according to the legend in panel~a). A horizontal black line and grey shading shows the inferred UV slope of \JGSzthirteenLA for a pure power-law (PL) continuum, while the horizontal dashed line indicates the upper limit obtained for the intrinsic power-law slope when including nebular continuum. Grey shading similarly indicates the inferred range on the top $\xi_\text{ion}$ axis. The horizontal dotted line shows the blackbody spectral slope in the Rayleigh-Jeans limit, $F_\lambda \propto \lambda^{-4}$.
	}
	\label{fig:Blackbody_spectra}
\end{figure*}

\subsubsection{Systemic redshift}
\label{sssec:Spectral_modelling_results_redshift}

We find the resulting systemic redshift to be marginally higher than in \citetalias{2025Natur.639..897W} as a result of the improved NIRSpec wavelength solution, which differs from previous results both in terms of the NIRSpec calibration files (\cref{ssec:Data_reduction}) and the additional empirical zero-point correction (\cref{sssec:Comparison_DEEP-JWST_OASIS_continuum_line}). This can be traced back most directly by the observed \Lya wavelength, here inferred to be $\lambda_\text{obs} = 1.7171_{-0.0037}^{+0.0040} \, \mathrm{\upmu m}$, compared to $\lambda_\text{obs} = 1.7084 \pm 0.0014 \, \mathrm{\upmu m}$ before. The same comparison also illustrates that the introduction of $\Delta p_i$ as a refined wavelength correction (\cref{ssec:Spectral_modelling_setup}) ensures that the systematic uncertainty of the wavelength solution is propagated into uncertainty estimates on all other quantities. Though degeneracy naturally exists between the two parameters $\Delta p_i$ (see \cref{app:Corner_plot}), the overall correction converges reasonably well as seen by the $\ssim2\sigma$ deviations away from $\Delta p_i = 0$, as would be preferred by the prior.

Finally, we do not recover clear detections of any rest-frame UV emission lines other than \Lya. The OASIS spectrum does show a modest peak at the expected location of the $\CIV \, \lambda \, 1548, 1551 \, \Angstrom$ doublet, for which we infer a tentative $3.0\sigma$ detection of $F_\text{CIV} = (11.0 \pm 3.7) \, \times 10^{-20} \, \mathrm{erg \, s^{-1} \, cm^{-2}}$ and $\text{EW}_\text{CIV} = 22.0 \pm 7.4 \, \Angstrom$. However, this signature is not confirmed in the JADES spectrum ($F_\text{CIV} = (2.8 \pm 3.7) \, \times 10^{-20} \, \mathrm{erg \, s^{-1} \, cm^{-2}}$), nor does its implied redshift match up with other emission-line signatures. Upper limits on the EWs of rest-frame UV features remain at the $20 \, \Angstrom$ level ($3\sigma$) even in the combined $56 \, \mathrm{h}$ JADES+OASIS PRISM spectrum.

We conclude that future spectroscopy will need to go substantially deeper to detect rest-frame UV features, which may be weakened as a result of a non-zero LyC escape fraction. Alternatively, medium-resolution spectroscopy as planned in the NIRSpec/G235M observations of JWST PID 9016 (PI: D.~Stark) may be more efficient at isolating narrow rest-UV lines from the underlying continuum emission, while a complementary strategy is to pursue more luminous lines at longer wavelengths, such as the $\OIII \, 88 \, \mathrm{\upmu m}$ that will be targeted in the upcoming ALMA programme 2025.1.00147.S (PI: J.~Witstok).

\section{Discussion}
\label{sec:Discussion}

Based on NIRCam imaging and the initial NIRSpec/PRISM spectroscopy obtained as part of JADES, \citetalias{2025Natur.639..897W} concluded that \JGSzthirteenLA hosts a compact ($r \lesssim 35 \, \mathrm{pc}$), powerful source of ionising radiation---a dense cluster of massive stars or, as also favoured by \citet{2026PASA...43...46C}, an AGN---responsible for driving the observed \Lya emission. Additionally, the efficient production and escape of LyC photons could also explain the likely presence of a modestly sized ionised bubble (with characteristic radius $R_\text{ion} \approx 0.2 \, \mathrm{pMpc}$ based on the model adopted in \cref{ssec:Spectral_modelling_results}). In the following sections, we will revisit and revise these arguments.

\subsection{The ionising source(s) and interstellar medium conditions within \JGSzthirteenLA}
\label{ssec:Ionising_source_and_ISM_conditions_within_GS-z13-1}

Leveraging the completed JADES and new OASIS observations, the \Lya line itself and the very blue continuum are clearly confirmed. Meanwhile, our renewed statistical analysis indicates that the data are consistent with what is expected from a fully neutral IGM without necessitating DLA absorption ($\Delta \text{BIC} = 10$ in favour of the model without DLA fit to the JADES and OASIS measurements). This resolves the potential issue of having to explain the superposition of DLA absorption and \Lya emission, although as discussed in \citetalias{2025Natur.639..897W}, this phenomenon has been reported in AGN \citep{2024MNRAS.532.4703W} and nearby star-forming galaxies \citep{2023ApJ...956...39H}. Moreover, in the absence of dust, the overall \Lya transmission could be aided to some degree by resonant scattering in neutral gas, whereby the \Lya photons would gain a systemic velocity offset that decreases their probability of subsequently being scattered out of the line of sight by neutral gas in the IGM.

However, it is difficult to completely rule out a scenario with DLA absorption, as suggested by the initial JADES spectrum presented in \citetalias{2025Natur.639..897W}. It is also conceivable that the observed difference in continuum turnovers between JADES and OASIS---perhaps seen most clearly in the notable residuals of the JADES spectrum around $\lambda_\text{obs} \approx 1.8 \, \mathrm{\upmu m}$ in \cref{fig:Model_spectra}c---is real. The small NIRSpec micro-shutters have been shown to cause `pseudo-slit losses' due to self-subtraction of spatially extended \Lya emission, although this effect is only expected to be relevant much closer to the line resonance \citep[$\lambda_\text{emit} \lesssim 1250 \, \Angstrom$;][]{2025MNRAS.542..128B}, and we do not see clear evidence of self-subtraction (\cref{app:Grating_spectra}). However, the two micro-shutter placements could preferentially probe the continuum from regions with different \HI column densities, which simulations indeed predict to vary significantly between sightlines even on spatial scales of $\lesssim 100 \, \mathrm{pc}$ \citep{2025arXiv251001315G}.

In any case, the constraints on the UV continuum shape again point towards \JGSzthirteenLA being a highly efficient producer of ionising radiation. Reaching a UV slope of $\beta_\text{UV} \approx -3$ is already challenging when the spectrum is dominated by O-type stars, the most massive and hottest stars in standard stellar models with effective surface temperatures of $T_\text{eff} = \num{40000} \, \mathrm{K}$ \citep[e.g.][]{2024MNRAS.529.4087T}. Moreover, as already discussed in \cref{ssec:Spectral_modelling_results}, the relative reddening by any nebular continuum emission requires an even bluer continuum intrinsically \citep[see also][]{2025OJAp....8E.104K}. This is illustrated in \cref{fig:Blackbody_spectra}, where we compare the observed UV slope with ideal blackbody curves and nebular continua for a range of electron temperatures and densities. At least to first order, a combination of these two components is expected to give rise to the observed continuum in \JGSzthirteenLA.

First considering the effective slope reached by pure blackbody emission, the qualitative comparison in \cref{fig:Blackbody_spectra}b suggests that the extremely blue continuum of \JGSzthirteenLA requires a source (or population of sources) reaching an (average) temperature of approximately $T \approx \num{50000} \, \mathrm{K}$. The effect is magnified when a nebular continuum is present, in which case even higher temperatures of $T \gtrsim \num{60000} \, \mathrm{K}$ are required. This regime is only reached by extreme objects typically representing a small fraction of stellar populations: for instance, Wolf-Rayet (WR) stars that have shed their outer envelopes through winds or eruptions \citep{2007ARA&A..45..177C}, stars stripped in binaries \citep[e.g.][]{2017A&A...608A..11G}, or very massive stars \citep[VMSs;][]{2022A&A...659A.163M}. Alternatively, as discussed in \citetalias{2025Natur.639..897W}, a truncated accretion disc can reach spectral slopes steeper than $\beta_\text{UV} = - 7/3 \approx -2.33$ \citep[the standard thin-disc model prediction;][]{1973A&A....24..337S}. Owing to the high effective temperature, all of these sources emit a large fraction of their light beyond the hydrogen ionisation edge, which indeed fits in with the required high $\xi_\text{ion}$ (\cref{ssec:Spectral_modelling_results}). Independently, the model constraints on $\xi_\text{ion}$ also translate to a blackbody temperature close to $T \approx \num{60000} \, \mathrm{K}$ (\cref{fig:Blackbody_spectra}b).

Apart from the \Lya line, the characteristics of \JGSzthirteenLA bear notable resemblance to UNCOVER-37126, a galaxy at $z = 10.3$ that is also similarly spatially compact \citep{2023MNRAS.524.5486A, 2024ApJ...977..250F}. Based on its blue UV continuum measured by NIRSpec ($\beta_\text{UV} \approx -2.9$) and the absence of strong emission lines covered by MIRI spectroscopy \citep[as well as ALMA observations;][]{2026OJAp....965333A}, \citet{2026A&A...711A.301M} recently inferred UNCOVER-37126 to have a near-unity LyC escape fraction. This implies an almost complete absence of the interstellar medium (ISM), attributed by \citeauthor{2026A&A...711A.301M} to extremely efficient star formation and/or strong feedback effects. However, excessive gas removal and LyC escape are unlikely to be the case in \JGSzthirteenLA, precisely because of the strong \Lya emission, while an episode of high LyC leakage in the recent past was argued in \citetalias{2025Natur.639..897W} to be disfavoured based on the very short recombination timescale at $z \approx 13$ ($x_\text{\HI} \approx 0.01$ after just $1 \, \mathrm{Myr}$). Another possibility, then, is a significant contribution from nearby galaxies to the establishment of such an ionised bubble, which will be discussed in more detail below.

\subsection{The environment of \JGSzthirteenLA: an early overdensity driving a large ionised bubble?}
\label{ssec:Environment}

Remarkably, the one other galaxy with confirmed \Lya emission at $z > 10$, GN-z11 in the GOODS-N field \citep{2023A&A...677A..88B, 2024A&A...687A.283S}, was revealed to have several nearby photometric candidates consistent with its spectroscopic redshift, $z = 10.6$ \citep{2023ApJ...952...74T}. Recently, another prominent galaxy overdensity candidate at $z \approx 10.5$ in GOODS-S was reported in \citet{2026ApJ..1006...18W}. While the candidate member galaxies at this point rely on photometric selection techniques, there is tentative evidence for enhanced \Lya transmission within this structure. The analysis by \citet{2026ApJ..1006...18W} tantalisingly reveals a radial trend in excess flux within the photometric band containing \Lya, suggesting that the \Lya transmission may be enhanced towards the centre of the overdensity. At an estimated $4\times$ enhancement in galaxy number density over the average field, this overdense region is indeed a prime candidate for hosting an early ionised bubble.
\begin{figure*}
	\centering
	\includegraphics[width=\linewidth]{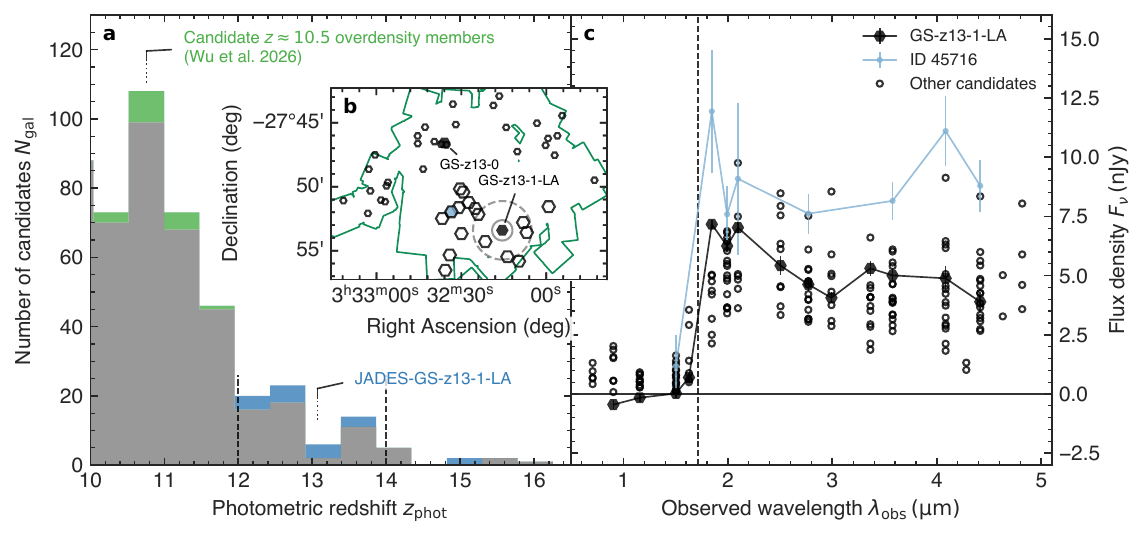}
	\caption{(a) Histogram of photometric redshifts from \citet{2026ApJ..1004..161H}. Members of the candidate overdensity identified by \citet{2026ApJ..1006...18W} are highlighted in green, those found in the vicinity of \JGSzthirteenLA in blue. Two $z \approx 15$ candidates enter the selection, since accounting for DLA absorption significantly decreases their estimated redshift (see text for details). (b) Spatial distribution of galaxy candidates found within the JADES NIRCam footprint (solid green line; cf. \cref{ssec:JWST_observing_programmes}) at $12 < z_\text{phot} < 14$, which are shown as small, open black hexagons. Larger hexagons mark the positions of JADES-GS-z13-0 \citep{2023NatAs...7..611R, 2023NatAs...7..622C, 2024ApJ...976..160H}, \JGSzthirteenLA (both filled black and annotated), and $17$ nearby galaxy candidates. The solid and dashed grey circles around \JGSzthirteenLA show the projected sizes of the ionised regions in our single-source and collective model, $R_\text{ion} \approx 0.2 \, \mathrm{pMpc}$ or $0.5 \, \mathrm{pMpc}$, respectively (\cref{ssec:Spectral_modelling_results}). (c) NIRCam photometry of \JGSzthirteenLA (black hexagons) and nearby candidates (open circles). A vertical dashed line indicates the observed \Lya wavelength in \JGSzthirteenLA, which falls close to the F182M medium-band filter where a tentative flux excess is seen in the brightest candidate (ID 45716, highlighted in blue).
	}
	\label{fig:Photometric_redshift_distribution}
\end{figure*}

In the following, we investigate whether there is any evidence for a similar overdensity at the redshift of \JGSzthirteenLA. We selected sources based on the catalogue of $z > 8$ galaxy candidates found in the GOODS-N and GOODS-S fields presented in \citet{2026ApJ..1004..161H}, which in turn are drawn from the latest JADES NIRCam DR5 \citep{2026arXiv260115954J, 2026arXiv260115956R}. Adopting similar criteria as in \citet{2024ApJ...962..124H, 2024ApJ...974...41H} and \citet{2025MNRAS.536...27W}, we considered photometric redshifts obtained by \citet{2026ApJ..1004..161H} using \textsc{eazy} \citep{2008ApJ...686.1503B}, where we selected sources in GOODS-S at $12 < z_\text{phot} < 14$, a conservatively broad redshift range that will be motivated further below. We furthermore required the redshift to be relatively well-constrained, $\Delta z_1/(1+z) < 0.25$, where $\Delta z_1$ is the difference between the \nth{16} and \nth{84} percentiles of the posterior redshift distribution. Finally, we narrow down the selection to sources whose \nth{84} (\nth{16}) percentile falls at $z > 12.8$ ($z < 13.5$), which acts to remove the spectroscopically confirmed sources JADES-GS-z12-0 at $z_\text{spec} = 12.48 \pm 0.01$ \citep[ID 96216;][]{2023NatAs...7..611R, 2023NatAs...7..622C, 2024A&A...689A.152D} and JADES-GS-z14-1 at $z_\text{spec} = 13.86_{-0.05}^{+0.04}$ \citep[ID 18044;][]{2024Natur.633..318C, 2025ApJ...992..212W}, respectively.

Especially at the highest redshift regime ($z \gtrsim 9$), however, photometric redshifts have regularly been found to be overestimated, at times significantly so \citep[e.g.][]{2023ApJ...949L..25F, 2023ApJ...952...74T, 2025ApJ...983L...2A, 2026OJAp....956033N, 2026ApJ..1002..134D}. This effect is, to a certain degree, expected from IGM damping-wing absorption \citep[e.g.][]{2024MNRAS.532.1646K, 2024MNRAS.531L..34K, 2025A&A...697A..89C, 2025ApJ...987...82H, 2026A&A...705A.114M}, but in some cases has also been shown to be exaggerated by extreme DLA absorption \citep[e.g.][]{2024ApJ...976..160H, 2024A&A...689A.152D, 2025ApJ...987L...2H, 2025A&A...696A..87C, 2026A&A...711A.298P, 2026OJAp....955261W}. To remedy this potential bias, in addition to choosing an upper redshift boundary of $z_\text{phot} = 14$, we considered the \citet{2025ApJ...983L...2A} prescription that was investigated in \citet{2026ApJ..1004..161H}.

For details of this prescription and its effectiveness in recovering photometric redshifts accurately and with reduced bias, we refer to \citet{2025ApJ...983L...2A} and \citet{2026ApJ..1004..161H}, but we briefly describe the method here. Specifically, the standard IGM prescription implemented in \textsc{eazy} \citep{2014MNRAS.442.1805I} is supplemented by a separate, empirically motivated DLA absorption component, which is calibrated on a sample of spectroscopically confirmed galaxies to be $N_\text{\HI} \approx 10^{21.5} \, \mathrm{cm^{-2}}$ at $z \gtrsim 8$. The effect of this prescription is very mild, bringing down photometric redshifts by only $\Delta z \approx 0.2$ on average \citep{2025ApJ...983L...2A}. To be conservative in our selection, we allow the redshift selections described above to be satisfied either under the default approach in \citet[; i.e. using the standard IGM treatment in \textsc{eazy}]{2026ApJ..1004..161H} or for redshifts obtained with the empirical \citet{2025ApJ...983L...2A} prescription. Photometric redshifts between both approaches are compared in \cref{app:Photometric_candidates}, which illustrates that its effects are mostly minor (relative to the typical uncertainties on $z_\text{phot}$), although notably two sources nominally at $z \sim 15$ are shifted down to solutions consistent with $z \sim 13$ (within uncertainties).

On the other hand, strong \Lya emission in principle can also cause photometric redshifts to be underestimated \citep[e.g.][]{2026ApJ..1006...18W}. Taking $\text{EW}_\text{\Lya} = 66 \Angstrom$ as in \JGSzthirteenLA as an example, the line effectively adds $0.09 \, \mathrm{\upmu m}$ of continuum bluewards of the true \Lya break, which decreases the photometric redshift by $\Delta z = (0.09 \, \mathrm{\upmu m}  / \lambda_\text{\Lya}) = 0.8$. While many \textsc{eazy} template sets include \Lya emission, additional systematic uncertainty on redshift is introduced when the true \Lya flux significantly deviates from the template, as also remarked by \citet{2024ApJ...970...31R} even before the spectroscopic confirmation of \JGSzthirteenLA.

Overall, we find the selection procedure to generally be robust against minor changes in the selection criteria. Interestingly, we retrieve $17$ candidates that are clustered within an angular separation of $\lesssim 6\arcmin$ from \JGSzthirteenLA. This translates to a projected distance of just over $1 \, \mathrm{pMpc}$, similar to the structure reported in \citet{2026ApJ..1006...18W}. By virtue of the initial selection in \citet{2026ApJ..1004..161H}, the best-fit photometric redshift solution is statistically preferred over any $z < 7$ solution ($\Delta \chi^2 > 4$), but we have furthermore verified for each of the $17$ candidates that filters bluewards of \Lya do not contain significant detections ($p > 0.1$ under the null hypothesis of zero flux in all filters). The NIRCam IDs, coordinates, redshifts, UV magnitudes and slopes of these candidates, as inferred by \citet{2026ApJ..1004..161H}, can be found in \cref{app:Photometric_candidates}.

The spatial distribution and NIRCam photometry of these candidates are shown in \cref{fig:Photometric_redshift_distribution}. Perhaps most strikingly, the brightest candidate (ID 45716) shows a tentative flux excess in the F182M medium-band filter, certainly compared to other candidates such as ID 54533 that show a much smoother break. This filter could contain the \Lya line if it falls slightly redwards compared to \JGSzthirteenLA \citepalias[for which no clear medium-band photometric boost is observed, since the line falls right between the F162M and F182M filters;][]{2025Natur.639..897W}. As noted in \citet{2026ApJ..1004..161H}, this bright galaxy candidate moreover shows an extended, complex morphology, including in F182M. Meanwhile, ID 11457 was already identified in \citet{2024ApJ...964...71H}, \citet{2024ApJ...970...31R}, and \citet{2025ApJ...992...63W} as a compact $z \approx 13$ galaxy candidate similar to \JGSzthirteenLA, if ever so slightly fainter.

Apart from the $17$ nearby photometric candidates, it is further worth mentioning that the spectroscopically confirmed galaxy JADES-GS-z13-0 \citep{2023NatAs...7..611R, 2023NatAs...7..622C, 2024ApJ...976..160H} is found at $8.2\arcmin$ from \JGSzthirteenLA ($1.7 \, \mathrm{pMpc}$). While the continuum break of this source initially suggested $z_\text{break} = 13.2$, as discussed above this may be overestimated by up to $\Delta z \approx 0.5$ due to the effects of DLA absorption \citep{2026OJAp....955261W}. Indeed, based on a deeper ($56 \, \mathrm{h}$) spectrum \citet{2024ApJ...976..160H} reported a more conservative redshift estimate of $z_\text{spec} = 13.13_{-0.13}^{+0.09}$ when taking into account potential DLA absorption, and tentative emission-line detections yielding $z_\text{spec} \approx 12.9$ have been found \citep{2024ApJ...976..160H, 2026A&A...711A.298P}, which would place GS-z13-0 close to the redshift of \JGSzthirteenLA.

We now turn to evaluate whether the presence of a larger ionised bubble, of the order of $R_\text{ion} \approx 0.5 \, \mathrm{pMpc}$ (\cref{ssec:Spectral_modelling_results}), is feasible. As a starting point, we adopt the best-fit Schechter function to the UV luminosity function (UVLF) derived by \citet{2025ApJ...992...63W} as broadly representative of the current literature consensus \citep[e.g.][]{2023ApJ...946L..13F, 2023ApJS..265....5H, 2024ApJ...960...56H, 2023ApJ...951L...1P, 2023ApJ...954L..46L, 2024ApJ...964...71H, 2024ApJ...970...31R, 2024ApJ...965..169A, 2024MNRAS.533.3222D}. This particular UVLF from \citet{2025ApJ...992...63W} was based on number counts of F150W dropouts, whose median redshift corresponds to $z = 12.8$. While the conversion of the UVLF into an ionising photon density $\dot{n}_\text{ion}$ hinges on several key assumptions, we aim to obtain a conservatively high estimate here. We integrate the UVLF down to an absolute UV magnitude of $M_\text{UV, lim} = -13 \, \mathrm{mag}$ ($M_\text{UV, lim} = -15 \, \mathrm{mag}$ would reduce $\dot{n}_\text{ion}$ by approximately $2\times$), consider an average ionising-photon production efficiency of $\xi_\mathrm{{ion}} = 10^{25.5} \, \mathrm{{Hz \, erg^{{-1}}}}$ (higher by a factor of several compared to \citealt{2024MNRAS.535.2998S} and \citealt{2025ApJ...981..134P}, though perhaps more appropriate for fainter systems; \citealt{2024Natur.626..975A}), and assume a relatively high average LyC escape fraction of $f_\text{esc, LyC} = 20\%$ \citep[e.g.][ and references therein]{2022ARA&A..60..121R}. These assumptions yield a value of $\dot{n}_\text{ion} \approx 2 \times 10^{50} \, \mathrm{s^{-1} \, cMpc^{-3}}$, yet even such a high rate falls short of ionising a spherical volume with radius $R_\text{ion} \approx 0.5 \, \mathrm{pMpc}$ within the entire lifetime of the Universe at $z = 13$.

Given the tentatively overdense environment of \JGSzthirteenLA, however, the ionised bubble growth could be sped up considerably: a $\gtrsim 3\times$ enhancement of the galaxy density compared to the average field, as deduced for the overdensity identified by \citet[; $4\times$ in that case]{2026ApJ..1006...18W}, could bring down the bubble formation timescale to the order of $100 \, \mathrm{Myr}$ or less. Indeed, the confirmed presence of \JGSzthirteenLA itself already implies its immediate surroundings are expected to be biased, given the high degree of clustering of galaxies particularly at $z \gtrsim 10$ \citep[e.g.][]{2023MNRAS.526L..47M, 2024ApJ...975..192G}. While we rely on small number statistics here, a first-order estimate suggests a number density enhancement of a factor of a few is certainly plausible. Over a $1 \, \mathrm{mag}$ range in $M_\text{UV}$ we would expect $N_\text{exp} \approx 0.4_{-0.3}^{+0.4}$ ($5_{-3}^{+5}$) galaxies within a sphere with a $0.5 \, \mathrm{pMpc}$ ($1 \, \mathrm{pMpc}$) radius, whereas we observe $N_\text{obs} = 5$ ($14$) candidates within consistent projected distances of \JGSzthirteenLA (including itself).

Altogether this leads us to conclude that it is feasible for \JGSzthirteenLA and its neighbouring galaxies to collectively inhabit a relatively large ionised bubble. However, this hinges on several optimistic assumptions: the ionising photon rate is derived using high estimates of $\xi_\text{ion}$ and $f_\text{esc, LyC}$, and the photometric candidates all require spectroscopic confirmation. Furthermore, we note that an overdense region is not only expected to linearly increase the production rate of ionising photons, but also to cause a local enhancement in the gas density \citep[e.g.][]{2024MNRAS.531L..34K, 2025MNRAS.544.2316K, 2026arXiv260819311Z}. This effect will inhibit the formation of an ionised region in a non-linear manner, since the recombination rate scales quadratically with density. Hence, it is unclear if an overdense environment at such an early stage of reionisation augments \Lya transmission or not. A critical test for the scenario with a galaxy overdensity will therefore be to search for additional \Lya emission lines in nearby sources such as the candidates identified in \cref{fig:Photometric_redshift_distribution}.

Either way, this does not take away from the deduction that \JGSzthirteenLA itself should be a prolific producer of ionising photons, a significant fraction of which may be escaping the galaxy, given the extremely blue continuum that allows little room for reprocessing into a redder nebular continuum (\cref{ssec:Ionising_source_and_ISM_conditions_within_GS-z13-1}). Whether or not a larger bubble is confirmed in the future, the conclusion therefore stands that early galaxies such as \JGSzthirteenLA can be extremely efficient ionising agents, and that reionisation was likely underway at $z \approx 13$.

\section{Conclusions}
\label{sec:Conclusions}

We have presented new JWST/NIRSpec spectroscopy of \JGSzthirteenLA. Our main findings can be summarised as follows:
\begin{itemize}
    \item We have combined new, deep NIRSpec/PRISM observations from the OASIS programme \citep{2026arXiv261041489L, 2026arXiv261041490L} with the now completed JADES observations, which with a total of $56 \, \mathrm{h}$ has tripled the total exposure time compared to \citetalias{2025Natur.639..897W}. This reaffirms the \Lya line detection and the blue UV continuum observed previously, although the OASIS spectrum suggests a less pronounced downturn of the continuum near \Lya. Deeper medium-resolution observations do not reveal a significant line detection, yet remain compatible with the PRISM measurements if the line is spectrally resolved.
    \item Detailed spectral modelling of the line and continuum emission generally supports the findings from \citetalias{2025Natur.639..897W}, except that the previously inferred DLA absorption with large \HI column density is statistically disfavoured with the addition of the new OASIS spectrum ($\Delta \text{BIC} = 10$). It is difficult to completely rule out the DLA solution, however, since the JADES and OASIS spectra could probe sightlines with truly different \HI column densities due to the difference in micro-shutter placement. A minor wavelength discrepancy in the blue NIRSpec coverage ($\lambda_\text{obs} \lesssim 2 \, \mathrm{\upmu m}$) can be explained by small remaining calibration systematics for spatially compact sources near the edges of micro-shutters, urging caution with PRISM-derived measurements in those cases.
    \item We investigated the implications of the measured \Lya line strength and steep continuum slope, which again point towards \JGSzthirteenLA hosting an ionising source with a high effective temperature ($T \gtrsim \num{60000} \, \mathrm{K}$), and allow at most a modest contribution from the nebular continuum.
    \item Combined, this implies that ionising photons may escape \JGSzthirteenLA at a sufficient rate to lead to the formation of a small ionised bubble with $R_\text{ion} \approx 0.2 \, \mathrm{pMpc}$. However, we emphasise that this represents a characteristic radius based on the model adopted in \cref{ssec:Spectral_modelling_results}, since the precise bubble size depends sensitively on model assumptions, most notably the prior on the \Lya velocity offset (\cref{sssec:Spectral_modelling_results_bubble}). Ruling out the presence of an ionised bubble, on the other hand, requires a fine-tuned scenario with a very red-skewed \Lya line profile that we deem unlikely.
    \item Based on the photometric candidates identified by \citet{2026ApJ..1004..161H}, we also identified $17$ nearby photometric candidates that together could plausibly establish a larger bubble ($R_\text{ion} \approx 0.5 \, \mathrm{pMpc}$). This could alleviate the required ionising photon production efficiency of \JGSzthirteenLA from $\xi_\mathrm{{ion}} \approx 10^{26.4} \, \mathrm{{Hz \, erg^{{-1}}}}$ down to $\approx 10^{25.9} \, \mathrm{{Hz \, erg^{{-1}}}}$, but it would equally require a notable overdensity of galaxies with highly efficient ionising capabilities.
\end{itemize}

\noindent We conclude that the new observations presented here confirm the overall picture of \JGSzthirteenLA as a remarkably powerful ionising source. Observed just $330 \, \mathrm{Myr}$ after the Big Bang, it provides compelling evidence to extend the timeline of cosmic reionisation towards perhaps unexpectedly early times. Upcoming observations, including a Cycle-4 JWST programme (PID 9016, PI: D.~Stark) obtaining deep NIRSpec/G235M spectroscopy and a Cycle-12 ALMA programme (2025.1.00147.S, PI: J.~Witstok) targeting the $\OIII \, 88 \, \mathrm{\upmu m}$ line, will help further unveil a more detailed picture of what may be the earliest stages of reionisation.

\begin{acknowledgements}
    We thank Andrea Ferrara, Joshua Cohon, and Charlotte Mason for insightful discussions. We furthermore thank the anonymous referee for helping to improve the manuscript. This work is based on observations made with the National Aeronautics and Space Administration (NASA)/European Space Agency (ESA)/Canadian Space Agency (CSA) JWST. The data were obtained from the Mikulski Archive for Space Telescopes at the STScI, which is operated by the Association of Universities for Research in Astronomy, Inc., under NASA contract NAS 5-03127 for JWST. These observations are associated with programmes 1180, 1210, 1287, 3215, and 5997. JW and AJC gratefully acknowledge support from the Cosmic Dawn Center through the DAWN Fellowship. The Cosmic Dawn Center (DAWN) is funded by the Danish National Research Foundation under grant No. 140. SC acknowledges support by European Union’s HE European Research Council (ERC) Starting Grant ``WINGS'' (Grant agreement No. 101040227). AJB acknowledges funding from the ``FirstGalaxies'' Advanced Grant from the ERC under the European Union’s Horizon 2020 research and innovation programme (No. 789056). FDE, GCJ, RM, and JS acknowledge support by the Science and Technology Facilities Council (STFC), by the ERC through Advanced Grant ``QUENCH'' (No. 695671), and by the UK Research and Innovation (UKRI) Frontier Research grant RISEandFALL. JMH, BER, BDJ, and CNAW are supported by the JWST/NIRCam Science Team contract to the University of Arizona, NAS5-02105, along with JWST programmes 3215 (JMH, BER, and BDJ) and 8544 (JMH). TJL gratefully acknowledges support from the Swiss National Science Foundation through a SNSF Mobility Fellowship and from the NASA/JWST programme OASIS (PID 5997). WMB gratefully acknowledges support from DARK via the DARK fellowship. This work was supported by a research grant (VIL54489) from VILLUM FONDEN. RM also acknowledges funding from a research professorship from the Royal Society. ST acknowledges support by the Royal Society Research Grant G125142. This work has relied on the following \textsc{python} packages: the \textsc{scipy} library \citep{2020NaMet..17..261V}, its packages \textsc{numpy} \citep{2011CSE....13b..22V} and \textsc{matplotlib} \citep{2007CSE.....9...90H}, \textsc{astropy} \citep{2013A&A...558A..33A, 2018AJ....156..123A}, \textsc{emcee} \citep{2013PASP..125..306F}, \textsc{forcepho}, and \textsc{(py)multinest} \citep{2009MNRAS.398.1601F, 2014A&A...564A.125B}.
\end{acknowledgements}

\bibliographystyle{aa}
\bibliography{aa60076-26.bib}

\begin{appendix}
\nolinenumbers
\onecolumn

\section{Two-dimensional NIRSpec/PRISM spectra and NIRSpec/G235M observations}
\label{app:Grating_spectra}

\begin{figure}[!htbp]
	\centering
	\sidecaption
	\includegraphics[width=12cm]{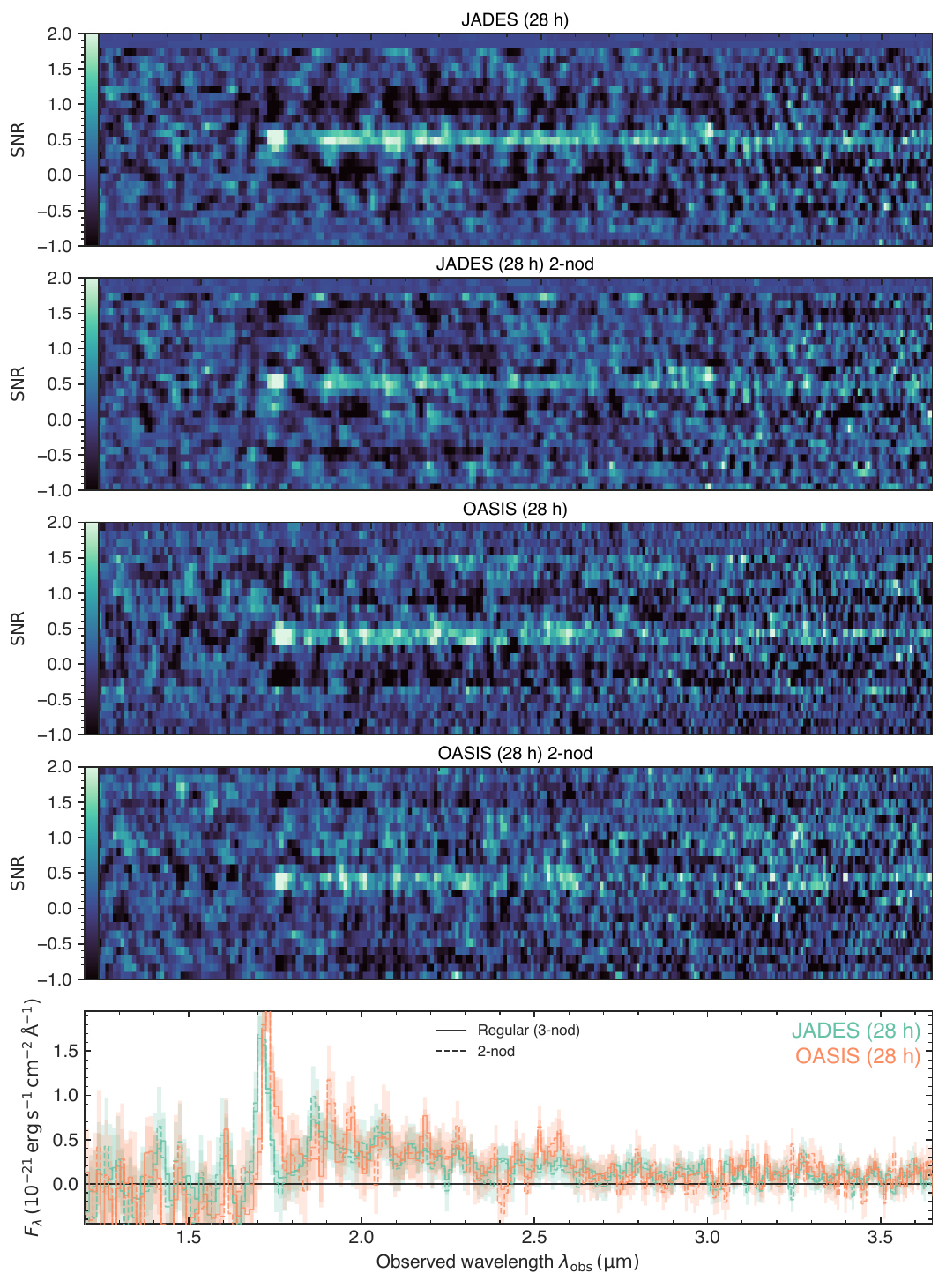}
	\caption{Two-dimensional spectra from JADES and OASIS, both in the default reduction where all three nod positions are considered (`3-nod'), and in the case where only the two outer nod positions are considered (`2-nod', leading to reduced SNR but avoiding potential self-subtraction). The line and continuum emission is observed to be spatially compact. The bottom panel further shows little difference between one-dimensional spectra extracted in the two reductions, suggesting that the effect of self-subtraction is limited.
	}
	\label{fig:2D_spectra}
\end{figure}

\begin{figure}[!htbp]
	\centering
	\includegraphics[width=\linewidth]{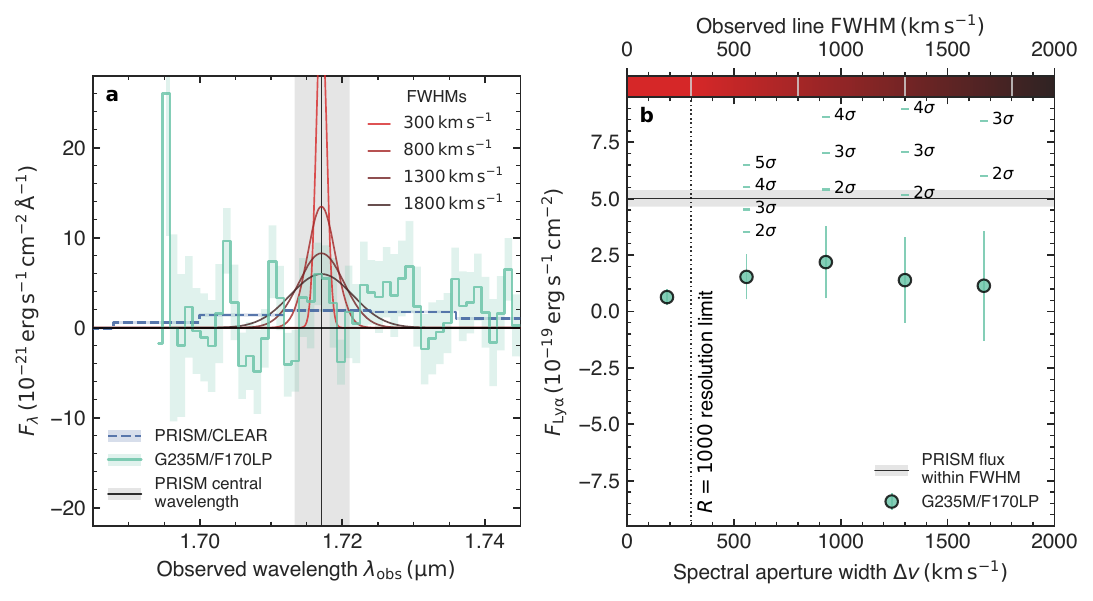}
	\caption{(a) Combined NIRSpec/G235M spectroscopy from JADES and the DSS survey as part of PID 3215 (\cref{ssec:JWST_observing_programmes}). Modelled \Lya emission-line profiles (solid curves) with matched flux and central wavelength (vertical black line) to the PRISM measurements are shown at different values of the FWHM according to legend and the grey vertical lines in the colour bar in panel~b, increasing from the $R = 1000$ resolution limit. (b) Inferred G235M \Lya flux in an incrementally widened spectral aperture (circles), none of which show a significant detection. The fraction of the PRISM line flux that would be contained within the FWHM of a Gaussian profile (76\%) is indicated by the horizontal black line and grey shading, while the $R = 1000$ resolution is shown by a vertical dotted line. If the \Lya line is intrinsically broad ($\text{FWHM} \gtrsim 600 \, \mathrm{km \, s^{-1}}$), the current measurements are not strongly in tension ($\lesssim 2\sigma$; cf. annotated significance levels).
	}
	\label{fig:Grating_spectra}
\end{figure}

In \cref{fig:2D_spectra}, the two-dimensional and one-dimensional NIRSpec/PRISM spectra are shown both in the default reduction where all three nod positions are considered, and in the case where only the two outer nod positions are considered (leading to reduced SNR but avoiding potential self-subtraction). In \cref{fig:Grating_spectra}, we show the combined NIRSpec/G235M spectra from JADES and the DSS survey as part of PID 3215 (\cref{ssec:JWST_observing_programmes}).


\clearpage

\section{Full spectral model posterior}
\label{app:Corner_plot}

\begin{figure}[!htbp]
	\centering
	\includegraphics[width=\linewidth]{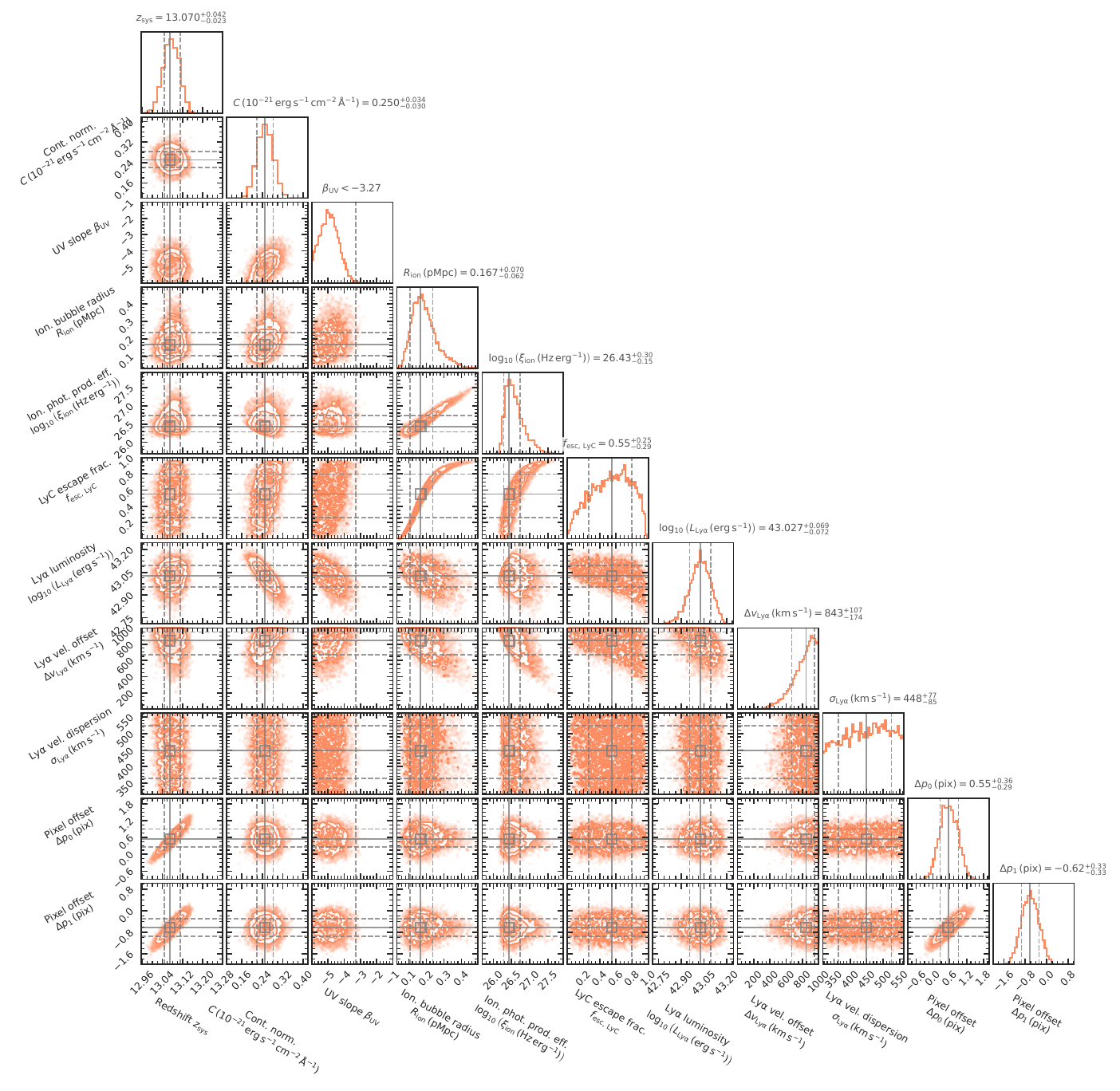}
	\caption{Posterior distributions obtained from spectral modelling of the PRISM spectrum of \JGSzthirteenLA (\cref{ssec:Spectral_modelling_setup,ssec:Spectral_modelling_results}). Individual panels show the inter-dependencies between the $9$ free parameters of the model, as well as the derived \Lya luminosity $L_\text{\Lya}$ (cf. \cref{tab:Model_results}).
	}
	\label{fig:Corner_plot}
\end{figure}

In \cref{fig:Corner_plot}, we show the full posterior distributions of the fiducial spectral model (\cref{ssec:Spectral_modelling_setup}).

\clearpage

\section{Photometric candidates}
\label{app:Photometric_candidates}

In \cref{tab:Photometric_candidates}, we list \JGSzthirteenLA along with the $17$ photometric candidates and $1$ spectroscopically confirmed source \citep[JADES-GS-z13-0;][]{2023NatAs...7..611R, 2023NatAs...7..622C, 2024ApJ...976..160H} found in its vicinity (\cref{ssec:Environment}).

\begingroup
    \setlength{\tabcolsep}{6pt}
    \renewcommand{\arraystretch}{1.5}
    \begin{table}[!htbp]
        \centering
        \footnotesize
        \caption{Photometric candidates identified in \citet{2024ApJ...970...31R} and \citet{2026ApJ..1004..161H} in the vicinity of \JGSzthirteenLA (\cref{ssec:Environment}).}
        \begin{tabular}{lllllllll}
            \toprule
            ID & $\text{RA} \, (\mathrm{deg})$ & $\text{Dec} \, (\mathrm{deg})$ & $z$ & $z_\text{DLA}$ & $M_\text{UV} \, (\mathrm{mag})$ & $\beta_\text{UV}$ & $\theta \, (\mathrm{arcmin})$ & $\theta \, (\mathrm{pMpc})$ \\
            \midrule
            41223 & 53.10013 & -27.86948 & $12.25_{-1.37}^{+1.28}$ & $12.11_{-1.75}^{+1.16}$ & $-17.71 \pm 0.13$ & $-1.8 \pm 0.5$ & $2.3$ & $0.5$ \\
            296077 & 53.033974 & -27.880013 & $12.33_{-1.05}^{+1.86}$ & $12.16_{-0.96}^{+1.55}$ & $-17.72 \pm 0.09$ & $-2.6 \pm 0.4$ & $1.7$ & $0.4$ \\
            354289 & 53.04021 & -27.93045 & $12.36_{-0.95}^{+1.03}$ & $12.06_{-0.71}^{+1.08}$ & $-18.00 \pm 0.13$ & $-2.9 \pm 0.4$ & $2.7$ & $0.6$ \\
            10346 & 53.12498 & -27.89461 & $12.41_{-1.12}^{+1.73}$ & $11.95_{-1.04}^{+1.54}$ & $-18.36 \pm 0.12$ & $-2.0 \pm 0.4$ & $3.2$ & $0.7$ \\
            508167 & 52.996464 & -27.85939 & $12.59_{-0.80}^{+0.97}$ & $11.83_{-0.82}^{+0.89}$ & $-17.97 \pm 0.13$ & $-2.3 \pm 0.4$ & $4.1$ & $0.9$ \\
            343416 & 53.057137 & -27.924593 & $12.69_{-1.27}^{+1.49}$ & $12.28_{-1.12}^{+1.29}$ & $-18.16 \pm 0.11$ & $-2.1 \pm 0.4$ & $2.1$ & $0.4$ \\
            377520 & 53.089848 & -27.904896 & $12.72_{-1.23}^{+1.26}$ & $12.37_{-1.09}^{+1.12}$ & $-18.05 \pm 0.10$ & $-2.4 \pm 0.4$ & $1.6$ & $0.3$ \\
            54533 & 53.129128 & -27.860743 & $12.78_{-0.37}^{+0.30}$ & $12.49_{-0.32}^{+0.61}$ & $-18.34 \pm 0.10$ & $-2.4 \pm 0.3$ & $3.8$ & $0.8$ \\
            52569 & 53.104687 & -27.86187 & $12.83_{-0.61}^{+0.73}$ & $12.66_{-0.44}^{+0.53}$ & $-17.76 \pm 0.09$ & $-2.4 \pm 0.3$ & $2.7$ & $0.6$ \\
            11457 & 53.02868 & -27.89301 & $13.04_{-0.40}^{+1.11}$ & $13.41_{-0.51}^{+0.58}$ & $-18.39 \pm 0.06$ & $-1.5 \pm 0.3$ & $1.9$ & $0.4$ \\
            13731\tablefootmark{$\ast$} & 53.06475 & -27.890234 & $13.07_{-0.02}^{+0.04}$\tablefootmark{$\ast\ast$} & -- & $-18.44 \pm 0.07$ & $-3.2 \pm 0.3$\tablefootmark{$\ast\ast$} & -- & -- \\
            84402 & 53.126827 & -27.836336 & $13.08_{-0.43}^{+0.76}$ & $12.53_{-0.30}^{+0.50}$ & $-17.73 \pm 0.21$ & $-2.5 \pm 0.7$ & $4.6$ & $1.0$ \\
            128771\tablefootmark{$\dagger$} & 53.149883 & -27.776503 & $13.13_{-0.13}^{+0.09}$\tablefootmark{$\ddagger$} & -- & $-18.55 \pm 0.04$ & $-2.7 \pm 0.1$\tablefootmark{$\ddagger$} & $8.2$ & $1.7$ \\
            45716 & 53.139744 & -27.866167 & $13.21_{-0.75}^{+0.79}$ & $12.19_{-0.59}^{+0.78}$ & $-18.75 \pm 0.07$ & $-1.8 \pm 0.4$ & $4.2$ & $0.9$ \\
            502974 & 53.14879 & -27.942274 & $13.43_{-1.00}^{+1.01}$ & $12.34_{-0.81}^{+0.96}$ & $-18.21 \pm 0.12$ & $-2.7 \pm 0.6$ & $5.4$ & $1.1$ \\
            535088 & 53.153336 & -27.874805 & $13.46_{-0.84}^{+0.71}$ & $12.42_{-0.68}^{+0.77}$ & $-17.70 \pm 0.15$ & $-2.0 \pm 1.0$ & $4.8$ & $1.0$ \\
            536773 & 53.142136 & -27.922363 & $13.56_{-1.47}^{+1.31}$ & $13.28_{-1.36}^{+1.23}$ & $-18.03 \pm 0.11$ & $-1.3 \pm 0.6$ & $4.5$ & $1.0$ \\
            184202 & 53.11405 & -27.854027 & $14.86_{-0.81}^{+0.67}$ & $13.81_{-0.60}^{+0.51}$ & $-18.90 \pm 0.09$ & $-1.8 \pm 0.4$ & $3.4$ & $0.7$ \\
            275184 & 53.1224 & -27.840427 & $14.94_{-0.73}^{+0.59}$ & $13.97_{-0.61}^{+0.51}$ & $-18.00 \pm 0.15$ & $-2.9 \pm 0.7$ & $4.3$ & $0.9$ \\
            \bottomrule
        \end{tabular}
        \tablefoot{
            Listed properties are the NIRCam ID, right ascension (RA), declination (Dec), photometric redshift both without and with the \citet{2025ApJ...983L...2A} DLA prescription ($z$ and $z_\text{DLA}$, respectively), absolute UV magnitude ($M_\text{UV}$), and UV slope ($\beta_\text{UV}$), as reported in the \citet{2026ApJ..1004..161H} catalogue (except for JADES-GS-z13-0 and \JGSzthirteenLA; see footnotes below). Also listed is the (projected) separation $\theta$ from \JGSzthirteenLA. \\
            \tablefoottext{$\ast$}{\JGSzthirteenLA.} \\
            \tablefoottext{$\ast\ast$}{Spectroscopic redshift and UV slope obtained in \cref{ssec:Spectral_modelling_results}.} \\
            \tablefoottext{$\dagger$}{JADES-GS-z13-0.} \\
            \tablefoottext{$\ddagger$}{Spectroscopic redshift (taking into account DLA absorption) and UV slope as inferred by \citet{2024ApJ...976..160H}.}
        }
        \label{tab:Photometric_candidates}
    \end{table}
\endgroup

\end{appendix}
\end{document}